\documentclass[aps,pra,reprint,10pt,a4paper]{revtex4-1}
\usepackage[utf8]{inputenc}
\usepackage[sc,osf]{mathpazo}
\usepackage{amsmath}
\usepackage[T1]{fontenc}
\usepackage{latexsym}
\usepackage{amssymb}
\usepackage[colorlinks=true,citecolor=blue,urlcolor=blue]{hyperref}
\usepackage{color}
\usepackage{graphics,epstopdf}
\usepackage{soul}
\usepackage[demo]{graphicx}
\usepackage{capt-of}
\usepackage{lipsum}
\usepackage{adjustbox}
\usepackage[normalem]{ulem}
\usepackage[table,xcdraw]{xcolor}
\begin{document}

\title{
Superiority in dense coding  through non-Markovian stochasticity}

\author{Abhishek Muhuri, Rivu Gupta, Srijon Ghosh, Aditi Sen(De)}

\affiliation{Harish-Chandra Research Institute,  A CI of Homi Bhabha National
Institute, Chhatnag Road, Jhunsi, Prayagraj - 211019, India}

\begin{abstract}

We investigate the distributed dense coding (DC) protocol, involving multiple senders and a single or two receivers under the influence of non-Markovian noise, acting on the encoded qubits transmitted from the senders to the receiver(s). We compare the effects of non-Markovianity on DC for both the dephasing and depolarising channels. In the case of dephasing channels, we illustrate that for some classes of states, high non-Markovian strength can eradicate the negative influence of noisy channels which is not observed for depolarizing noise. Furthermore, we incorporate randomness into the noise models by replacing the Pauli matrices with random unitaries and demonstrate the constructive impact of stochastic noise models on the quenched averaged dense coding capacity. Interestingly, we report that the detrimental effect of non-Markovian depolarising channels in the DC protocol can be eliminated when randomness is added to the channel. 
\end{abstract}

\maketitle

\section{Introduction}
\label{sec:intro}

Entanglement, the nonlocal resource shared with distant partners, is the most inherent difference between a conventional classical information-sharing protocol and its quantum version \cite{HoroRMP}. Basic protocols like dense coding \cite{bennettwiesner}, teleportation \cite{teleoriginal}, key distribution \cite{EkertCrypto}, and one-way quantum computation \cite{Raussendorf_PRL_2001} take advantage of this salient feature to surpass the classical limit imposed by unentangled resource states. In particular, quantum dense coding (DC) 
is the process of transmitting classical information, encoded in quantum states, across a long distance using an entangled state as the channel \cite{bennettwiesner}. 
The successful implementation of the DC protocol consists of three steps \cite{Bose, Bowen_PRA_2001, Liu, Ziman, Hiroshima, Bruss, DCCamader, Laurenza_PRR_2020} - (i) encoding at the sender's end, (ii) sending the encoded part through a quantum channel connecting the sender and the receiver, and (iii) decoding the information via measurements at the receiver's site. Moreover, the transmission of classical information has also been experimentally demonstrated over reasonably large separation of the parties with physical systems like photons \cite{Mattle_PRL_1996, Shimizu_PRA_1999, Mizuno_PRA_2005, Pan_RMP_2012, Northup_NP_2014, Barreiro_NP_2014, Krenn_PNAS_2016}, and trapped ions \cite{Fang_PRA_2000, Libfried_RMP_2003, Vandersypen_RMP_2005, Yang_JOPB_2007}.

The dense coding capacity (DCC) of a shared resource state, which is characterized as the maximum amount of classical information that can be sent, is used to quantitatively assess the performance of the shared state in the DC protocol. It can be shown that not all entangled states are beneficial for DC \cite{Gross}.  In particular, a quantum state is said to be densecodeable if and only if it can provide a capacity greater than what can be achieved through purely classical means \cite{Hausladen_PRA_1996,Bowen_PRA_2001, Horodecki_arxiv_2001}. 
If the encoded parts are transmitted over a noiseless channel and the Holevo bound  \cite{Holevo1973, Holevo-1998} is used during the decoding of the information, the compact version of DCC can be determined analytically by maximizing over unitary encoding for any number of senders and a single receiver.
When two receivers decoding the message via local operations and classical communication (LOCC) are involved in DC, the exact dense coding capacity is not known although the upper bound was provided \cite{Aditi} by employing local Holevo-like bound \cite{Localaccess, horodecki2004, ghosh2005}, which can be improved either by increasing the set of operations used for encoding \cite{Beran_PRA_2008, Shadman2011, Shadman_PRA_2012} or through pre-processing \cite{Gupta_PRA_2021}. 
Different kinds of DC schemes like port-based dense coding protocol \cite{Wu_PRA_2006,Zhengfeng_PRA_2006, Bourdon_PRA_2008, Tsai_OC_2010, Srivastava} and probabilistic dense coding have also been proposed \cite{Hao_PLA_2000,Feng_PRA_2006, Kogler_QIP_2017}. 


Since isolated systems cannot be prepared,  the systems under consideration are always in contact with the environment which, in general, is responsible for the decay of quantum properties like quantum correlations. Environmental noise is often classified into two categories -- Markovian \cite{nielsenchuang}, which has no memory effect during evolution, and non-Markovian \cite{Petruccione, Daffer_PRA_2004, rivas2012, Shrikant_PRA_2018, lidar2019}, that retains memories of earlier stages of the evolution and influences later noise processes.  It was found that quantum correlations, by nature, are fragile against noise injected by the environment, e.g., under local dephasing noise, entanglement suddenly dies away -- a phenomenon known as entanglement sudden death \cite{Ting_PRL_2004, Todd_PRA_2004, Ficek_PRA_2006, almeida2007, yu2009, Mazzola_PRA_2009}  which, in turn, affects all the information processing tasks. Specifically,  the DC protocol can be affected by noise in two ways -- first, when the resource states are shared, which has already been considered in the derivation of DCC \cite{Bose, Bowen_PRA_2001, Horodecki_arxiv_2001, Ziman, Bruss, DCCamader} and second, when encoded qubits are sent to the receiver(s) for which DCC affected by certain kinds of Markovian noise has  also been addressed \cite{Quek_PRA_2010, Shadman_NJP_2010, Shadman2011, Shadman_PRA_2012, Shadman2013, Das_PRA_2014, Das_PRA_2015, Mirmasoudi_JPA_2018}(cf. \cite{Liu_EPL_2016} for non-Markovian noise with a single sender-receiver pair).

In this work,  we examine the impact of both Markovian and non-Markovian noises on the multipartite dense coding protocol involving multiple senders and one or two receivers.
Our aim is to determine whether the non-Markovian nature of the noise has any favorable influence on the transmission of classical information, since the collapse and subsequent revival-like phenomena were observed in the instance of correlation measures \cite{Mazzola_PRA_2009, Rivas_PRL_2010, PMazzola_PRA_2010, Haikka_PRA_2013, LoFranco_2017, Karpat_2017, Gupta_PRA_2022}. 
Specifically, we demonstrate that when the shared states are the generalized Greenberger-Horne-Zeilinger (gGHZ) \cite{Greenberger_arXiv_2007,Lomonaco_arXiv_2004} and generalized W states (gW) \cite{Dur_PRA_2000,Aditi2003}, the DCC can be enhanced in the presence of high dephasing noise and high non-Markovianity compared to the Markovian dephasing noise, thereby eliminating the detrimental effects of noise on the protocol. Interestingly, we observe that for depolarizing non-Markovian noise, such an increment is absent.
Furthermore, the increase of DCC under non-Markovian noise is more pronounced in the case involving only two senders compared to the case with three senders.



On a different front, the inherently stochastic nature of quantum theory allows for the existence of randomness which is a thriving area of research in the context of random unitaries and circuits. Various studies have been devoted to the dynamics of entanglement, such as the properties of many-body systems \cite{Chan_PRL_2018, von_PRX_2018}, and to demonstrate operator spreading \cite{Nahum_PRX_2018} by random unitaries \cite{Nahum_PRX_2017, Zhou_PRB_2019, Li_arXiv_2022}.
It is quite unnatural to anticipate that the noise affecting resources will be of a specific sort, such as dephasing or depolarizing noise. Instead, a certain type of noise with some fluctuations, that can be represented by random unitaries selected from a Gaussian distribution with a fixed standard deviation around the Pauli matrix, can replicate a more realistic environment.
We perform quenched averaging of DCC over several such realizations of the choices of randomness injected on the noisy channels.
We manifest that for a fixed amount of non-Markovianity, the quenched averaged DCC and its upper bound for two receivers increase with the increase of disorder strength when the shared states are the gGHZ and the gW. Specifically, we highlight the increment of the averaged DCC  achieved under random non-Markovian depolarizing noise over the same under Markovian noise without any disorder.


Our paper is arranged in the following manner. In Sec. \ref{sec:pre}, we introduce random channels and certain quantities required for in-depth analysis of the noisy dense coding protocol. The positive impact of non-Markovianity on DCC is presented in Sec. \ref{sec:non-marko} while Sec. \ref{sec:random_channels}  demonstrates DCC under the random noise models. Finally, we summarise our observations in Sec. \ref{sec:conclusion}.

\section{Quality factors to assess the performance of DC under noise}
\label{sec:pre}
We lay down the framework required for the analysis of our findings in this section.  \textcolor{black}{We first present a brief outline of the noisy dense coding protocol involving an arbitrary number of senders and a single as well as two receivers. We then discuss how to construct random non-Markovian channels. Finally, quantities that capture the critical behavior of the noise strength for obtaining quantum advantage are introduced.}  

\subsection{Noisy Dense Coding Capacity }
\label{app:noisy_dcc}

\textcolor{black}{The dense coding capacity (DCC) quantifies the amount of classical information that can be transmitted with the help of a shared quantum state, known as the resource state. A second quantum channel is also required, through which the senders transfer their qubits to the receiver(s) after encoding the message. When the encoded information is sent through a noiseless channel, the DCC for an $(N+1)$-party state, $\rho_{\mathcal{S}_1...\mathcal{S}_{N}\mathcal{R}}$, shared between $N$ senders, $\mathcal{S}_i$ $(i = 1, 2, 3, \dots, N)$ and a single receiver, $\mathcal{R}$, reads \cite{Bruss,DCCamader}
\begin{eqnarray}
       \nonumber  C^1(\rho_{\mathcal{S}_1...\mathcal{S}_{N}\mathcal{R}}) = \text{max} && [\log_2 d_{\mathcal{S}_1...\mathcal{S}_{N}},\log_2 d_{\mathcal{S}_1...\mathcal{S}_{N}} + \\
       && S(\rho_\mathcal{R}) - S(\rho_{\mathcal{S}_1...\mathcal{S}_{N}\mathcal{R}})].
        \label{eq:DCC_3-1}
\end{eqnarray}
Here, $\log_2 d_{\mathcal{S}_1 \dots \mathcal{S}_N} = \log_2 d_{\mathcal{S}_1} d_{\mathcal{S}_2} \dots d_{\mathcal{S}_N}$ corresponds to the capacity without quantum advantage, referred to as the classical bound, with $d_{\mathcal{S}_i} (i = 1, 2, \dots, N)$ being the dimension of the subsystem of the sender, $\mathcal{S}_i$. $S(\rho) = -\text{Tr}(\rho \log_2 \rho)$ is the von Neumann entropy and $\rho_\mathcal{R}$ is the reduced subsystem at the receiver's end, obtained by tracing out the senders' part of the original state, i.e., $\rho_\mathcal{R} = \text{Tr}_{\mathcal{S}_1...\mathcal{S}_{N}}\rho_{\mathcal{S}_1...\mathcal{S}_{N}\mathcal{R}}$. A resource state is said to be densecodeable when $S(\rho_\mathcal{R}) - S(\rho_{\mathcal{S}_1...\mathcal{S}_{N}\mathcal{R}}) > 0$, signifying quantum advantage over the classical bound. Note that Eq. \eqref{eq:DCC_3-1} is obtained by considering that the encoding performed by the senders is through unitary operations (cf. \cite{Horodecki_2012}). We refer to this protocol as $N \mathcal{S} - 1 \mathcal{R}$.}

\textcolor{black}{When the channel through which the senders transmit their qubits to the receiver is noisy, the multipartite DCC has been shown to be \cite{Shadman_NJP_2010, Das_PRA_2014}
\begin{eqnarray}
       \nonumber C^1_{\text{noise}}(\rho_{\mathcal{S}_1...\mathcal{S}_{N}\mathcal{R}}) = \text{max} && [\log_2 d_{\mathcal{S}_1...\mathcal{S}_{N}},\log_2 d_{\mathcal{S}_1...\mathcal{S}_{N}} + \\
       && S(\rho_R) - S(\tilde{\rho})],
       \label{eq:DCC_3-1_noisy}
\end{eqnarray}
where
\begin{eqnarray}
   \nonumber  \tilde{\rho} = \Lambda(&& (U^{\min}_{\mathcal{S}_1}\otimes...\otimes U^{\min}_{\mathcal{S}_{N}} \otimes \mathbb{I}_\mathcal{R}) \rho_{\mathcal{S}_1...\mathcal{S}_{N}\mathcal{R}} \\
    &&(U^{\text{min} \dagger }_{\mathcal{S}_1} \otimes...\otimes U^{\text{min} \dagger }_{\mathcal{S}_{N}} \otimes \mathbb{I}_\mathcal{R})).
    \label{eq:rho_tilde}
\end{eqnarray}
Here $\Lambda$ is a completely positive trace preserving (CPTP) map denoting the noisy channel and $U^{\min}_{\mathcal{S}_i}$, a local unitary applied by the sender $i$, such that the von-Neumann entropy in the last term of Eq. \eqref{eq:DCC_3-1_noisy} is minimized. If the noisy channel $\Lambda$ is covariant, which means that it commutes with a complete set of orthogonal unitary operators, $\{W_{i}\}$ \cite{Shadman_NJP_2010, Shadman2013}, we have
\begin{equation}
    \Lambda(W_i \rho W^\dagger_i) = W_i \Lambda(\rho) W^\dagger_i.
\end{equation}
The noisy dense coding capacity with covariant noise reduces to
\cite{Shadman_NJP_2010, Das_PRA_2014}
\begin{eqnarray}
       \nonumber C^1_{\text{noise}}(\rho_{\mathcal{S}_1...\mathcal{S}_{N}\mathcal{R}}) = \text{max} && [\log_2 d_{\mathcal{S}_1...\mathcal{S}_{N}},\log_2 d_{\mathcal{S}_1...\mathcal{S}_{N}} + \\
       && S(\rho_\mathcal{R}) - S(\Lambda(\rho))].
       \label{eq:DCC_3-1_noisy_covariant}
\end{eqnarray}
Note that the simplification occurs since the noisy channel $\Lambda$,  commutes with the entropy-minimizing unitaries $U^{\min}_{\mathcal{S}_i}$ and the von-Neumann entropy is invariant under local unitary operators. A paradigmatic example of such a channel is the depolarising channel while the dephasing channel is not so. Thus, in order to estimate the DCC of a resource state in the presence of a noisy channel that is not covariant, the minimization over the unitaries $U^{\min}_{\mathcal{S}_i}$ has to be performed. In this work, we will consider that the senders' nodes are affected individually by both the dephasing and depolarising noises, after encoding.}


\subsubsection*{Distributed Super Dense Coding Capacity  in a noisy environment}
\label{subsubsec:noisy_dcc-2R}
\textcolor{black}{Let us consider that the state, $\rho_{\mathcal{S}_1...\mathcal{S}_{N}\mathcal{R}_1 \mathcal{R}_2}$, is shared between $N$ senders and two receivers such that the first $r$ senders, $\mathcal{S}_1,...,\mathcal{S}_r$, communicate their classical message to $\mathcal{R}_1$, while the remaining senders, $\mathcal{S}_{r+1},...,\mathcal{S}_{N}$, do so to $\mathcal{R}_2$.  Local operations and classical communication (LOCC) between two receivers  are allowed for decoding, and we call this scheme $N \mathcal{S}-2 \mathcal{R}$. Instead of the Holevo bound \cite{Holevo1973} which is used to derive the DCC for a single receiver, a Holevo-like bound for LOCC \cite{Localaccess} can be applied to obtain the distributed DCC. Since it is known that this bound cannot be achieved asymptotically, 
we obtain only the upper bound on the dense coding capacity which reads as \cite{DCCamader}
\begin{eqnarray}
       \nonumber  B^{2}(\rho_{\mathcal{S}_1...\mathcal{S}_{N}\mathcal{R}_1 \mathcal{R}_2} &&) = \text{max}  [\log_2 d_{\mathcal{S}_1...\mathcal{S}_{N}}, \log_2 d_{\mathcal{S}_1...\mathcal{S}_{N}} + \\
       && S(\rho_{\mathcal{R}_1}) + S(\rho_{\mathcal{R}_2}) - \text{max}_{\substack{x = 1,2}}S(\xi^x)], ~~~~~~
       \label{eq:DCC_2R}
\end{eqnarray}
where $\rho_{R_i}$ is the subsystem of the  receiver $i (i = 1,2)$, obtained by tracing out all the senders and the other receiver. Here, 
      \( \xi^1 = \text{Tr}_{\mathcal{S}_{r+1}..\mathcal{S}_{N}\mathcal{R}_2} \rho_{\mathcal{S}_1...\mathcal{S}_{N}\mathcal{R}_1 \mathcal{R}_2}\), ~~~ and
       \( \xi^2 = \text{Tr}_{\mathcal{S}_{1}..\mathcal{S}_{r}\mathcal{R}_1} \rho_{\mathcal{S}_1...\mathcal{S}_{N}\mathcal{R}_1 \mathcal{R}_2}\), 
where the former is obtained by tracing out the second group of senders $(\mathcal{S}_{r+1}, \dots, \mathcal{S}_N)$ and a receiver $\mathcal{R}_{2}$. Similarly, we obtain $\xi^2$.
Let us now suppose that the channels through which the two sets of senders transfer their qubits to the respective receivers are noisy, represented by $\Lambda$. The upper bound on the DCC alters as \cite{Das_PRA_2015}
\begin{eqnarray}
      \nonumber  B^{2}_{\text{noise}}(\rho_{\mathcal{S}_1...\mathcal{S}_{N}\mathcal{R}_1 \mathcal{R}_2}&&) =  \text{max}  [\log_2 d_{\mathcal{S}_1...\mathcal{S}_{N}}, \log_2 d_{\mathcal{S}_1...\mathcal{S}_{N}}  \\
       && \nonumber  + S(\rho_{\mathcal{R}_1}) + S(\rho_{\mathcal{R}_2}) - \text{max}_{\substack{x = 1,2}}S(\tilde{\xi}^x)].\\
       \label{eq:DCC_2R_noisy}
\end{eqnarray}
Here, $\tilde{\xi}^x$ again represents the  reduced subsystems of the corresponding senders-receiver pairs as
\begin{eqnarray}
\nonumber  \tilde{\xi}^1 = \nonumber\\ \nonumber \text{Tr}_{\mathcal{S}_{r+1}..\mathcal{S}_{N}\mathcal{R}_2} (\Lambda(&&(U^{\min}_{\mathcal{S}_1}\otimes...\otimes U^{\min}_{\mathcal{S}_{N}} \otimes \mathbb{I}_{\mathcal{R}_1 \mathcal{R}_2}) 
 \rho_{\mathcal{S}_1...\mathcal{S}_{N}\mathcal{R}_1 \mathcal{R}_2} \times \\
&& \nonumber (U^{\text{min} \dagger }_{\mathcal{S}_1}\otimes...\otimes U^{\text{min} \dagger }_{\mathcal{S}_{N}} \otimes \mathbb{I}_{\mathcal{R}_1 \mathcal{R}_2}))), \\ \\ \nonumber  
\label{eq:sender-receiver1_noisy}
\end{eqnarray}
and similarly \(\tilde{\xi}^2\). 
Like in the single receiver scenario, local unitaries $U^{\min}_{\mathcal{S}_i}$s are applied in order to minimize the entropy in the last term of Eq. \eqref{eq:DCC_2R_noisy}. Note that $U^{\min}_{\mathcal{S}_1}\otimes...\otimes U^{\min}_{\mathcal{S}_{r}}$ and $U^{\min}_{\mathcal{S}_{r+1}}\otimes...\otimes U^{\min}_{\mathcal{S}_{N}}$ independently minimize $S(\tilde{\xi}^1)$ and $S(\tilde{\xi}^2)$ respectively. As in Eq. \eqref{eq:DCC_3-1_noisy_covariant}, the covariant channels lead to a capacity similar to Eq. \eqref{eq:DCC_2R_noisy}, with a modification in $\tilde{\xi}^x$.}

\subsection{Action of random quantum channels}
\label{subsec:random_noise}

Exemplary noise models considered typically in the literature include
the dephasing, depolarising, amplitude damping channels \cite{Preskill}  (see Appendix \ref{subsec:noise_channel} for Kraus representation of the dephasing and depolarizing channels). However, in reality, they can seldom be realized accurately according to their Kraus representation involving the Pauli matrices. During the dense coding protocol, the noise acting on the encoded qubits at the senders' side may be quite different from the Pauli noise that characterizes some channels like dephasing and depolarizing. To address such a situation, local noise models based on random unitary operators are considered,  representing the noise that actually affects each qubit sent to the receiver(s). An arbitrary two-dimensional unitary matrix $U$ is parameterized by four variables as
\begin{equation}
     U = e^{i\phi}
    \begin{pmatrix}
    e^{i\omega/2} & 0 \\
    0 & e^{-i\omega/2}
    \end{pmatrix}
    \begin{pmatrix}
    \cos\theta/2 & -\sin\theta/2\\
    \sin\theta/2 & \cos\theta/2
    \end{pmatrix}
    \begin{pmatrix}
    e^{i\delta} & 0\\
    0 & e^{-i\delta}
    \end{pmatrix}.
    \label{eq:random_unitary}
\end{equation}
For brevity, we set $\phi = 0$, since it only contributes to an overall phase. Here, specific values of $\omega$, $\theta$, and $\delta$ lead to  the well-known Pauli matrices. The values of the three parameters which characterize the Pauli matrices (up to a global phase) are given by $\sigma_x: \omega = 2\pi, \theta = \pi, \delta = \pi$; $\sigma_y: \omega = 3\pi, \theta = \pi, \delta = \pi$; and $\sigma_z: \omega = 2\pi, \theta = 0, \delta = 3\pi$.
Note that the values of $\omega, \theta$, and $\delta$, mentioned above,  are not unique for a given Pauli matrix. We design the random noisy channels by replacing  $\sigma_i$s in their Kraus representation by random unitary matrices given in Eq. \eqref{eq:random_unitary}. The parameters of the random unitary $U_i$ corresponding to a given $\sigma_i (i = x, y, z)$ are chosen from a Gaussian distribution with the corresponding mean, \(\mu\)  and a fixed standard deviation, say $\epsilon$, denoted by \(\mathbb{G}(\mu,  \epsilon)\). For example, the random unitary $U_x$ corresponding to $\sigma_x$ has its parameters $\omega, \theta$, and $\delta$ chosen from  Gaussian distributions, $\mathbb{G}(2\pi,\epsilon), \mathbb{G}(\pi,\epsilon)$ and $\mathbb{G}(\pi,\epsilon)$ respectively, which in turn, quantifies the fluctuation around $\sigma_x$ during implementation. \textcolor{black}{Note that, when we characterize a random noise model, we incorporate the fluctuation in the standard Pauli matrices, keeping the framework of the dephasing and depolarising noise models intact. Precisely,  we use the same Kraus operators as standard noise models but with the Pauli matrices replaced by random unitaries $U$, each of which is a function of three parameters $\omega, \theta$, and $\delta$ chosen from a Gaussian distribution of the fixed mean which is the desired value   and a given standard deviation. Note that the desired  Pauli matrices are recovered with vanishing standard deviation.
This construction ensures that the random noise still remains a CPTP map and is physically realizable. Thus, the only modification in the Kraus operators is that they are defined by random unitaries instead of Pauli matrices (the coefficients remain the same as in the original definition of the dephasing and depolarising noise).} Hence, we can redefine the Kraus operators for the random non-Markovian dephasing  and depolarising  channels  respectively as  
\begin{eqnarray}
		&& \kappa_{I}^{dph} =\sqrt{[1-\alpha p](1-p)} \mathbb{I}, \kappa_{z}^{dph} = \sqrt{[1+\alpha(1-p)]p}U_z, \nonumber 
		\label{eq:random_deph}
		\end{eqnarray}
	with \( 0 \leq p \leq 1/2\)	and
	\begin{eqnarray}
	&& \kappa_{I}^{dp} =\sqrt{[1-3\alpha p](1-p)} \mathbb{I},  \kappa_{i}^{dp} = \sqrt{\frac{[1+3\alpha(1-p)]p}{3}}U_{i}, \nonumber\\
		  \label{eq:random_depo}
\end{eqnarray}
with $i = x, y$ and $z$. \textcolor{black}{The degree of non-Markovianity is denoted by $\alpha$, where a higher value indicates more backflow of information from the environment into the system, while the quantity $p$ represents the strength of noise acting on the system, i.e., an increase in $p$ implies that a greater amount of noise is affecting the system. In the Markovian limit (i.e., with $\alpha = 0$), the noise parameter $p$ varies as $0 \leq p \leq 0.5$ for the dephasing channel and as $0 \leq p \leq 1$ in the case of the depolarising channel. The allowed range of $p$ in the non-Markovian regime is the same as the Markovian one for the dephasing channel, whereas for finite non-Markovian strength ($\alpha > 0$) in the depolarising channel,  $p \in (0, \frac{1}{3 \alpha})$ to ensure that the Kraus operators for the depolarising channel remain positive and the channel remains a CPTP map.} Note that, unlike the dephasing channel, the depolarising one is covariant even in the presence of non-Markovianity. However, even a small deviation from the usual Pauli matrices makes the channel incovariant.

\textcolor{black}{\textbf{Remark}.  By replacing the Pauli matrices with random unitaries, the noise that affects the protocol becomes random, but since the Kraus operators retain their original form, the nature of the noise remains the same. For example, the action of the dephasing channel is to leave the state (on which it acts) unaffected with some probability, $p$ or change its phase by acting $\sigma_z$ with probability $(1 - p)$. In a similar way, for the random ``noisy'' dephasing channel, the action of the noise is to either keep the state unaffected with a certain probability $(p)$ or to change it by acting on it with a unitary close to the $\sigma_z$ operator (with probability $1 - p$). A similar argument holds for the noisy depolarising channel, where noise acts on the state from all directions $x, y, z$ but the action is quantified by random unitaries instead of the original Pauli matrices. Therefore, even in the presence of random unitaries, we refer to the noise models as ``noisy dephasing'' and ``noisy depolarising'' channels, since the characters of the noise remain unchanged through the coefficients of the Kraus operators, i.e., ``how" the noise disturbing the system does not change, but its eventual effect does not correspond to that of the well-known Pauli noise models.}

Below, we estimate $U^{\min}$ which is required to optimize the DCC in the presence of dephasing noise for a class of shared states.\\

\textbf{Proposition $\mathbf{1}$}. \textit{For the dense coding protocol involving $N$ senders and a single receiver sharing an $(N+1)$- qubit gGHZ state affected by local dephasing noise, the optimizing unitaries $U^{\min}$ corresponding to each sender are proportional to the identity operator}.\\

\textit{Proof}. \textcolor{black}{The $(N+1)$-qubit gGHZ state is given by $|gGHZ\rangle^{N + 1} = x |0\rangle^{\otimes N + 1} + \sqrt{1 - x^2}|1\rangle^{\otimes N + 1}$. The senders perform unitaries $U^{\text{min}}_{\mathcal{S}_j}(\omega_j, \theta_j, \delta_j)$ minimizing $S(\tilde{\rho})$ in the dense coding capacity $C^{1}_{max}(\rho_{s_{1}\ldots s_{N}R})$ (see Eq. (\ref{eq:DCC_3-1_noisy})), before being affected by the noise $\Lambda$. For each sender, $\mathcal{S}_j$, the two-dimensional unitary $U^{\text{min}}_{\mathcal{S}_j}$ can be parameterized (upto an overall phase) by three parameters $\omega_j, \theta_j$ and $\delta_j$. The noisy state $\Lambda(\tilde{\rho}_{gGHZ})$ has $2^{N + 1}$  eigenvalues, all of which are functions of $\{x, \alpha, p, \theta_j, \omega_j, \delta_j\}$ with $\alpha$ and $p$ being the strength of non-Markovianity and noise respectively. The analytical form of the eigenvalues required to compute \(S(\Lambda(\tilde{\rho}_{gGHZ}))\) is too complicated to present in the paper, but close observation reveals that they can be written as
\begin{eqnarray}
    \mu_k &=& e^{i \sum_{j = 0}^{N - 1}(\omega_j + \delta_j)}\sqrt{e^{-2 i \sum_{j = 0}^{N - 1}(\omega_j + \delta_j)} f_k(x, \alpha, p, \theta_j)} \nonumber\\
    &=& \sqrt{f_k(x, \alpha, p, \theta_j)},
    \label{eq:response_th1}
\end{eqnarray}
where $k = 1, 2, \cdots, 2^{N + 1}$ identifies the  eigenvalues. Note that the functional form of $f_k$ may be different for different eigenvalues and, the only condition that $f_k$ must satisfy is that $\sum_k \sqrt{f_k(x, \alpha, p, \theta_j)} = 1$, to ensure normalization. Even though the form of $f_k$ is complicated, Eq. \eqref{eq:response_th1} indicates that the minimization of $S(\Lambda(\tilde{\rho}_{gGHZ}))$ needs to be performed only over the variables $\theta_j$ of each $U^{\text{min}}_{\mathcal{S}_j}$, and not over $\omega_{j}$ and $\delta_{j}$. Note that this function $f$ does not have any physical significance, it only helps us to write the eigenvalues of the resulting state in a compact form. Numerical minimization of $S(\Lambda(\tilde{\rho}_{gGHZ}))$ over $\theta_j$ for $N$ up to $10$ reveals that the minimum  occurs at $\theta_{j_{\text{opt}}} \approx n \pi ~ \forall ~ j$. Thus, $\cos \theta_{j_{\text{opt}}} \approx \pm 1$ and $\sin\theta_{j_{\text{opt}}} \approx 0$ implying that the minimizing unitaries are proportional to the identity operator, i.e., $U^{\text{min}}_{\mathcal{S}_j} \propto \mathbb{I}$ (where $\mathbb{I}$ is the $2 \times 2$ identity operator). Hence the proof. $\hfill \blacksquare$}  \\
\textbf{Remark}. Note that if the shared state is an arbitrary $N$-qubit state, $U^{\min}$ may not simplify to the identity operator.

\subsection{Critical noise strengths}
\label{subsec:quant}
Let us define certain physical quantities that can help us analyze the effect of noise on the DCC. These quantities are introduced to capture the overall behavior that emerges in the DC protocol affected by decoherence and imperfections. In the presence of noise, one can expect the following changes in any indicator $\mathcal{Q}$ quantifying the protocol's performance, which is the dense coding capacity in this case. 
\begin{enumerate}
    \item Typically, $\mathcal{Q}$ decreases with the increase of the noise strength $p$ and vanishes either at a finite $p$ or asymptotically. Hence, we define the minimum value of the noise strength, referred to as the critical noise strength, and denoted by $p_c$, at which the DCC or quenched averaged DCC (which will be defined later for random noise) collapses to its classical limit for a fixed value of $\alpha$ as
    \begin{equation}
        p_c = \min_p[p | C^1_{\text{noise}}~\text{or}~B^2_{\text{noise}} = \log_2(d_{\mathcal{S}_1} \dots d_{\mathcal{S}_N})].
        \label{eq:pc}
    \end{equation}
    A lower value of $p_c$ indicates that the state involved in the protocol is more susceptible to noise and vice versa.
    
    \item In the presence of non-Markovian noise, $\mathcal{Q}$, in general, revives after collapse due to the backflow of information. For example, we know that it is the case for entanglement \cite{Rivas_PRL_2010, LoFranco_2017, Karpat_2017, Gupta_PRA_2022}. Motivated by this picture, the minimum value of $p$, at which the capacity first goes beyond its classical limit after collapse can be called the revival strength of noise - denoted by $p_r$, given by
    \begin{equation}
        p_r = \min_p[p \geq p_c|C^1_{\text{noise}}~\text{or}~B^2_{\text{noise}} > \log_2(d_{\mathcal{S}_1} \dots d_{\mathcal{S}_N})].
        \label{eq:pr}
    \end{equation}
    The constructive effect of the noise, if induced, is more prominent when the value of $p_r$ is low which indicates that the capacity revives faster.
    \item Moreover, it is possible to identify a range of the noise parameters, in which non-Markovianity helps to overcome the detrimental effect that occurred due to  Markovian noise, thereby illustrating the constructive effect of non-Markovianity. To compare the Markovian and non-Markovian noise scenarios, we define the quantity  manifesting the advantage furnished by non-Markovianity, denoted by $p_a$, as
    \begin{eqnarray}
      \nonumber   p_a = \min_p[p| && C^1_{\text{noise}}(\alpha>0)~ >   C^1_{\text{noise}}(\alpha=0)~\\
      && \text{or} ~~ B^2_{\text{noise}}(\alpha>0) > B^2_{\text{noise}}(\alpha=0)].
      \label{eq:pa}
    \end{eqnarray}
\end{enumerate}
We will investigate the patterns of the above quantities depending on the noise models and shared resource states, in the succeeding sections.
Note, however, that typically, to obtain a quantum advantage in the dense coding protocol, a high amount of entanglement is required and so it is not apriori guaranteed that the feature of revival that is observed for entanglement \cite{Gupta_PRA_2022}, can also be apparent for DCC. The above quantities  $p_c, p_r$, and $p_a$ highlight the different impacts of noise on the dense coding protocol. \textcolor{black}{But first, let us briefly explain how numerical optimization is employed to calculate the DCC in the presence of noise.}

\subsection{Numerical optimization method}
\label{subsec:nlopt}
\textcolor{black}{The quantity being minimized is $S(\Lambda((U^{\text{min}}_{\mathcal{S}_1} \otimes \cdots \otimes U^{\text{min}}_{\mathcal{S}_N} \otimes \mathbb{I}_{\mathcal{R}} )\rho_{\mathcal{S}_1 \cdots \mathcal{S}_N \mathcal{R}} ( U^{\text{min}\dagger}_{\mathcal{S}_1} \otimes \cdots \otimes U^{\text{min} \dagger}_{\mathcal{S}_N} \otimes \mathbb{I}_{\mathcal{R}})))$. Since each minimizing two-dimensional unitary $U^{\text{min}}_{\mathcal{S}_i}$ can be parameterized by three variables $\omega_i, \theta_i$, and $\delta_i$, the entropy function involves optimization over the $3N$ quantities $\{\omega_i, \theta_i, \delta_i\}_{i = 1}^{N}$ corresponding to the $N$ senders. The ranges of the variables are set as $0 \leq \omega_i, \delta_i \leq 2 \pi$ and $0 \leq \theta \leq \pi$. We use the Improved Stochastic Ranking Evolution Strategy (ISRES) algorithm based on the method described in Ref. \cite{Runarsson_IEEE_2005} to perform non-linear optimization using the NLOPT library \cite{nloptcpp} in C++. The evolution strategy involves two steps - mutation rule (with a log-normal step-size update and exponential smoothing) and differential variation. Since our optimization problem does not involve any non-linear constraint, the objective function itself determines the fitness ranking. The optimization algorithm is executed $10^{5}$ times in order to locate the global minimum. We use the aforementioned algorithm since it has the heuristics to escape local extrema present within the variable range. The convergence is set to $10^{-5}$ in the NLOPT routine, which guarantees that the minimum of the entropy function is correct up to the fifth decimal place. We first use the algorithm to find $U^{\text{min}}_{\mathcal{S}_i}$ for the depolarizing channel, for which the minimizing unitaries should reduce to the identity for each sender (since the noise is covariant). We verify that this is indeed the case, i.e., $U^{\text{min}}_{\mathcal{S}_i} = \mathbb{I}$ to ensure that the ISRES algorithm is suited for our purpose before applying it for other types of noise.}

\section{Dense Coding  influenced by non-Markovian noise}
\label{sec:non-marko}

We now investigate the response of non-Markovian dephasing and depolarising noise on the dense coding scheme. The noise acts on each channel that carries the encoded qubits from the senders' side to the receiver(s). Specifically, we aim to find out whether non-Markovian noise in the channel can have a constructive impact on the capacity or not. If yes, we are interested in identifying a range of parameters in the channel where such an effect is predominant.

Before going to a protocol with an arbitrary number of senders and a single or two receivers, let us consider the simplest bipartite scenario involving a single sender and a single receiver. One can check that $C^1_{\text{noise}} (\alpha >0) > C^1_{\text{noise}} (\alpha = 0)$ for high values of the noise parameter, $p$, and the non-Markovianity strength $\alpha$, when the shared state is $|\phi^+\rangle_{\mathcal{S} \mathcal{R}}$ (see Appendix. \ref{sec:bell_dcc}). It will be interesting to find whether such an advantage persists for arbitrary shared multipartite states in the $N$ senders - $1$ receiver and $N$ senders - $2$ receivers regimes. In this respect, note that it was shown that the relation between multipartite entanglement and DCC  differs from the bipartite domain
\cite{Das_PRA_2015} and hence the effects of noise observed in the case of DCC with two parties may not hold in the multipartite domain which is indeed the case as shown in the succeeding sections.

\subsection{Noisy dense coding between arbitrary senders and a single receiver}
\label{subsec:nm_dense_n-sender}

In the multipartite domain, a natural choice of the resource state is an $(N+1)$-party generalized GHZ (gGHZ) state, given by $|gGHZ\rangle^{N+1} = x |0\rangle^{\otimes N+1} + \sqrt{1-x^2} | 1\rangle^{\otimes N+1}$, shared between $N$ senders and a single receiver. Such a choice is due to the fact that in the noiseless scenario, $|gGHZ\rangle^{N+1}$ with $x = \sqrt{1/2}$ provides the maximum DCC, i.e., $\log_2 d_{\mathcal{S}_1} \dots d_{\mathcal{S}_N} + 1$ while for other values of $x$, the capacity reads as $\log_2 d_{\mathcal{S}_1} \dots d_{\mathcal{S}_N}  + H(\{x^2,1-x^2\})$ where $H(\{p_i\}) = - \sum_i p_i \log_2 p_i$ represents the Shannon entropy corresponding to the single-site reduced density matrix at the receiver's end having eigenvalues, $\{x^2, 1 - x^2\}$.

{\it Impacts of dephasing non-Markovian channel.} In the presence of both Markovian and non-Markovian noise $(\alpha \neq 0)$,  the dephasing channel is not covariant, which implies that the minimization over $U^{\min}$ corresponding to each of the $N$ senders has to be performed. However, such optimization is hard to perform analytically and so we will resort to numerical optimization in obtaining the exact trends of the DCC by varying the noise parameter and the strength of non-Markovianity.

Before proceeding further, we present a lower bound on the DCC in the presence of the non-Markovian dephasing channel.\\
\textbf{Theorem $\mathbf{2}$}. \textit{The lower bound on the DCC, obtained under the non-Markovian dephasing channel, provides an advantage over its Markovian counterpart, for a certain range of the noise parameter when an $(N+1)$- qubit gGHZ state is shared between $N$ senders and a single receiver.} \\

\textit{Proof}.  As stated before,  the DCC without noise for the shared gGHZ state, $|gGHZ\rangle^{N+1}$, with $N$ senders and a single receiver  can be obtained. In the presence of noise, motivated by  Proposition $1$, let us assume that $U^{\min} = \mathbb{I}_{\mathcal{S}_1} \otimes \mathbb{I}_{\mathcal{S}_2} \otimes \dots \otimes \mathbb{I}_{\mathcal{S}_N}$ (with $\mathbb{I}$ being the $2 \times 2$ identity matrix), which leads to a lower bound on the actual DCC. In this case, the capacity reduces to $C^1_{\text{noise}} = \log_2 d_{\mathcal{S}_1} \dots d_{\mathcal{S}_N} + H(\{x^2,1-x^2\}) - S(\Lambda(\rho^{N+1}))$ where $\rho^{N+1} = |gGHZ\rangle^{N+1} \langle gGHZ|$ with $\Lambda$ being the non-Markovian dephasing channel. Notice first, that under the action of the dephasing channel (both Markovian and non-Markovian), the coefficients in the density matrix only get modified but no additional coefficients appear beyond those of the pure state. This implies that there are only two eigenvalues, given by 
\begin{eqnarray}
     &&  \frac{1}{2} \Big(1 \pm \sqrt{1 - 4 (-1 + (1 - 2p)^{2N})x^2(x^2-1)}\Big),~~\text{and}\nonumber \\ 
    && \frac{1}{2} \Big(1 \pm \sqrt{1 - 4(-1 + (1 - 2p + 2 (p-1)p\alpha)^{2N})x^2(x^2-1)}\Big), \nonumber \\
    \label{eq:proof_ev_nonmarko_rho}
\end{eqnarray}
for the Markovian and non-Markovian channels respectively. 
Since both $S(\rho_\mathcal{R})$ and $S(\Lambda(\rho^{N+1}))$ are bi-variate functions, it is clear that the state is dense codeable when the smaller eigenvalue of $S(\Lambda(\rho^{N+1}))$ is lower than that of $S(\rho_{\mathcal{R}})$, thereby making $S(\rho_\mathcal{R}) \geq S(\Lambda(\rho^{N+1}))$. Let us consider that the eigenvalues of $\rho_\mathcal{R}$ are such that $x^2 < (1-x^2)$. In the case of the Markovian channel, it is observed that the lower eigenvalue of $\Lambda(\rho^{N+1})$ 
is less than $x^2$, when $p \leq 1/2$. Thus for the Markovian dephasing channel, the gGHZ state always gives a quantum advantage in the dense coding protocol. \\
Repeating the same exercise with the lower eigenvalue in Eq. \eqref{eq:proof_ev_nonmarko_rho}, we find that the non-Markovian dephasing noise allows for dense codeability of the gGHZ state when $p \neq p_c = (1 + \alpha - \sqrt{1 + \alpha^2})/2\alpha$, which is the noise strength at which the capacity reduces to its classical limit. Thus a higher value of $\alpha$ decreases the noise strength at which the state ceases to be dense codeable. 
However, comparing $S(\Lambda_{M}(\rho^{N+1}))$ and $S(\Lambda_{NM}(\rho^{N+1}))$, with the subscript being the types of noise in the channel, we find that when  $p \geq p_a = (2 + \alpha - \sqrt{4 + \alpha^2})/2 \alpha$,  $S(\Lambda_{M}(\rho^{N+1})) \geq S(\Lambda_{NM}(\rho^{N+1}))$, and hence the non-Markovian channel furnishes a higher dense coding capacity than the Markovian one. Since $p_a \geq p_c$, the non-Markovian advantage is guaranteed for $ [(2 + \alpha - \sqrt{4 + \alpha^2})/2 \alpha, 0.5] $, and a higher value of $\alpha$  indicates a greater range of improvement due to non-Markovianity since $p_a$ decreases monotonically with $\alpha$. Hence the proof. $\hfill \blacksquare$\\
\begin{table*}[]
	\caption{Critical noise strength $p_c$, revival noise strength $p_r$ and noise strength, $p_a$, for which advantage is furnished due to non-Markovianity for the GHZ and the W states affected by non-Markovian dephasing noise. } 
	\begin{tabular}{|c|crrr|cr|crr|}
\hline
$\alpha$                  & \multicolumn{4}{c|}{$p_c$}                                                                                                                                                                                        & \multicolumn{2}{c|}{$p_r$}                                                                              & \multicolumn{3}{c|}{$p_a$}                                                                                                                                   \\ \hline
                          & \multicolumn{1}{c|}{}                              & \multicolumn{1}{c|}{}                              & \multicolumn{1}{c|}{}                              & \multicolumn{1}{c|}{}                              & \multicolumn{1}{c|}{}                              & \multicolumn{1}{c|}{}                              & \multicolumn{1}{c|}{}                              & \multicolumn{1}{c|}{}                              & \multicolumn{1}{c|}{}                              \\ \hline
                          & \multicolumn{2}{c|}{GHZ}                                                                                & \multicolumn{2}{c|}{W}                                                                                  & \multicolumn{2}{c|}{GHZ}                                                                                & \multicolumn{2}{c|}{GHZ}                                                                                & \multicolumn{1}{c|}{W}                             \\ \hline
                          & \multicolumn{1}{c|}{}                              & \multicolumn{1}{c|}{}                              & \multicolumn{1}{c|}{}                              & \multicolumn{1}{c|}{}                              & \multicolumn{1}{c|}{}                              & \multicolumn{1}{c|}{}                              & \multicolumn{1}{c|}{}                              & \multicolumn{1}{c|}{}                              & \multicolumn{1}{c|}{}                              \\ \hline
                          & \multicolumn{1}{c|}{$2 \mathcal{S}-1 \mathcal{R}$} & \multicolumn{1}{c|}{$3 \mathcal{S}-1 \mathcal{R}$} & \multicolumn{1}{c|}{$2 \mathcal{S}-1 \mathcal{R}$} & \multicolumn{1}{c|}{$3 \mathcal{S}-1 \mathcal{R}$} & \multicolumn{1}{c|}{$2 \mathcal{S}-1 \mathcal{R}$} & \multicolumn{1}{c|}{$3 \mathcal{S}-1 \mathcal{R}$} & \multicolumn{1}{c|}{$2 \mathcal{S}-1 \mathcal{R}$} & \multicolumn{1}{c|}{$3 \mathcal{S}-1 \mathcal{R}$} & \multicolumn{1}{c|}{$2 \mathcal{S}-2 \mathcal{R}$} \\ \hline
\multicolumn{1}{|l|}{}    & \multicolumn{1}{l|}{}                              & \multicolumn{1}{l|}{}                              & \multicolumn{1}{l|}{}                              & \multicolumn{1}{l|}{}                              & \multicolumn{1}{l|}{}                              & \multicolumn{1}{l|}{}                              & \multicolumn{1}{l|}{}                              & \multicolumn{1}{l|}{}                              & \multicolumn{1}{l|}{}                              \\ \hline
\multicolumn{1}{|r|}{0}   & \multicolumn{1}{r|}{0.48}                          & \multicolumn{1}{r|}{0.42}                          & \multicolumn{1}{r|}{0.13}                          & 0.07                                               & \multicolumn{1}{r|}{}                              &                                                    & \multicolumn{1}{r|}{}                              & \multicolumn{1}{r|}{}                              &                                                    \\ \hline
\multicolumn{1}{|r|}{0.3} & \multicolumn{1}{r|}{0.41}                          & \multicolumn{1}{r|}{0.35}                          & \multicolumn{1}{r|}{0.1}                           & 0.06                                               & \multicolumn{1}{r|}{0.46}                          &                                                    & \multicolumn{1}{r|}{0.48}                          & \multicolumn{1}{r|}{}                              &                                                    \\ \hline
\multicolumn{1}{|r|}{0.5} & \multicolumn{1}{r|}{0.36}                          & \multicolumn{1}{r|}{0.31}                          & \multicolumn{1}{r|}{0.09}                          & 0.05                                               & \multicolumn{1}{r|}{0.41}                          & 0.47                                               & \multicolumn{1}{r|}{0.44}                          & \multicolumn{1}{r|}{0.47}                          & 0.45                                               \\ \hline
\multicolumn{1}{|r|}{0.7} & \multicolumn{1}{r|}{0.33}                          & \multicolumn{1}{r|}{0.28}                          & \multicolumn{1}{r|}{0.08}                          & 0.05                                               & \multicolumn{1}{r|}{0.37}                          & 0.42                                               & \multicolumn{1}{r|}{0.42}                          & \multicolumn{1}{r|}{0.42}                          & 0.41                                               \\ \hline
\multicolumn{1}{|r|}{0.9} & \multicolumn{1}{r|}{0.29}                          & \multicolumn{1}{r|}{0.25}                          & \multicolumn{1}{r|}{0.07}                          & 0.04                                               & \multicolumn{1}{r|}{0.33}                          & 0.38                                               & \multicolumn{1}{r|}{0.4}                           & \multicolumn{1}{r|}{0.4}                           & 0.39                                               \\ \hline
\end{tabular}
	\label{table_nm_deph}
\end{table*}
\textbf{Remark}. Although the depolarizing non-Markovian channel is covariant,  the depolarizing channel acting on the senders' part of the gGHZ state changes the state space drastically which, in turn,  increases the number of non-zero eigenvalues with the increase of $N$,  and the proof along the lines of the dephasing channel does not go through. However, numerical simulations in the case of the depolarizing channel become easy owing to its covariant property.\\
\begin{figure}[!ht]
		\centering
		\includegraphics[width=\linewidth]{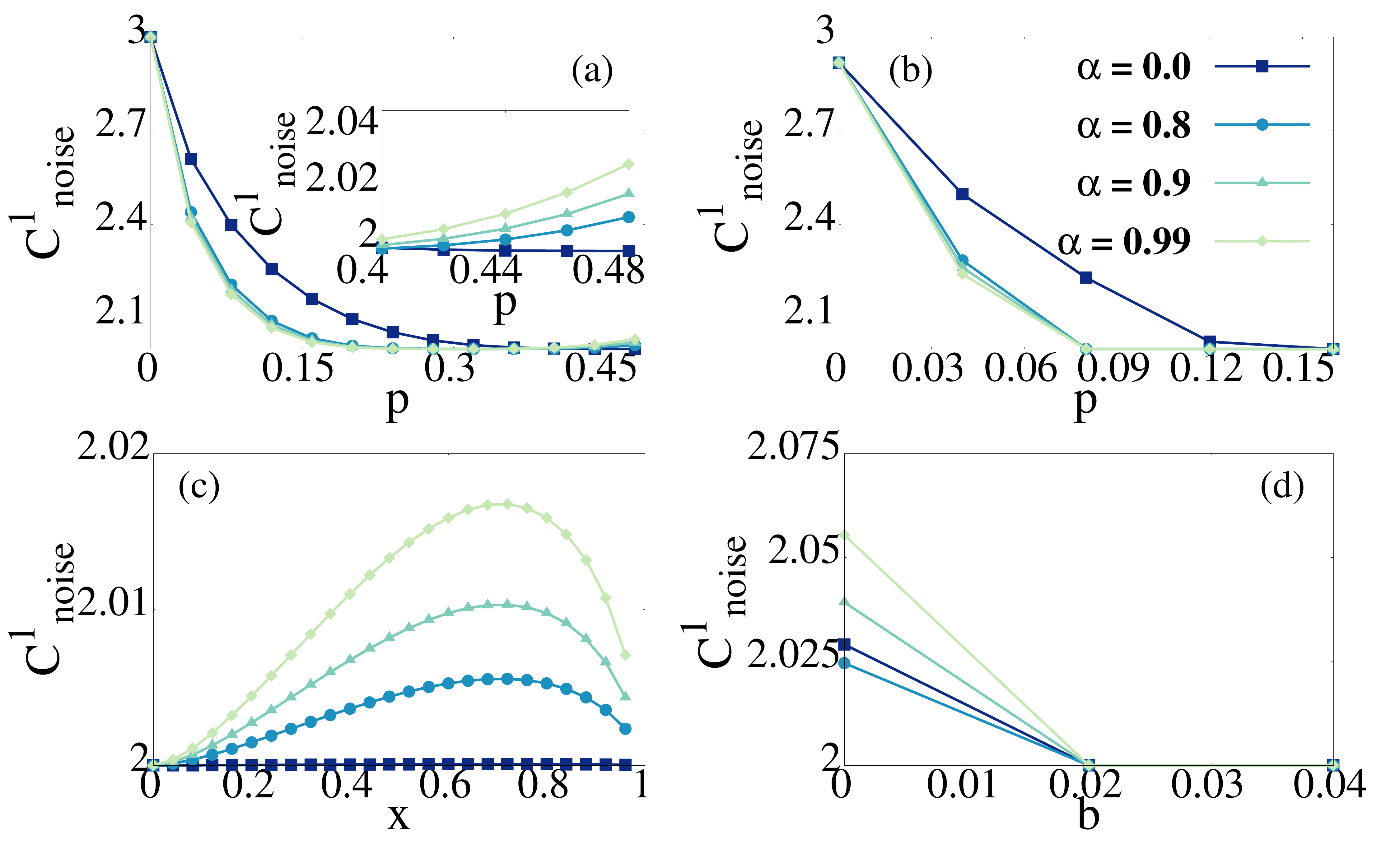}
		\caption{\textcolor{black}{(Color online.) {\bf Two senders and a single receiver, \(2 \mathcal{S} - 1\mathcal{R}\) case under non-Markovian dephasing noise.}
 (a) $C^1_{\text{noise}}$ (ordinate) against the non-Markovian dephasing noise parameter, $p$ (abscissa) for different non-Markovianity, \(\alpha\), when the shared resource state is the three-qubit GHZ which is affected by dephasing noise after encoding. The inset in (a) highlights the constructive effect of non-Markovianity at high noise strength on the GHZ state; (b) A similar study is performed for the shared $W$ state. (c) and (d) illustrate DCC (ordinate) of three-qubit $|gGHZ\rangle^{3}$ and $|W_{1/2}\rangle^{3}$ states respectively against the state parameters $x$ and $b$ (abscissa) respectively for $p = 0.4$. The different non-Markovianity parameters are represented from dark (blue) to light (green) as $\alpha = 0.0$ (squares), $\alpha = 0.8$ (circles), $\alpha = 0.9$ (triangles) and $\alpha = 0.99$ (diamonds) respectively. The \(x\)-axis is dimensionless whereas the \(y\)-axis is in bits.}}
	\label{fig:2s-1r-deph}
\end{figure}

\subsubsection{Dense coding under non-Markovian noise}
\label{subsubsec:dcc_non-Marko_deph}

Let us now concentrate on the exact DCC  affected by non-Markovian dephasing noise by performing numerical optimization. Let us now consider two exemplary sets of pure states as the shared resource - the gGHZ state and the generalized W state, given by $|gW\rangle^{N+1} = \sum_i b_i \mathcal{P} [|0\rangle^{\otimes N} |1\rangle]$ where the $b_i$s are chosen to be real, and $\mathcal{P}$ denotes the permutation operator which permutes the vector $|1\rangle$ in different positions. The gW state for three-qubits \cite{Agrawal_PRA_2006, ROY_PLA_2018}, reduces to $|gW\rangle^3 = \sqrt{a}|001\rangle + \sqrt{b} |010\rangle + \sqrt{1 - a - b}|100\rangle$ where $a, b \in \mathbb{R}$ and  $a + b \leq 1$ which, for $a = 1/2$,  exhibits perfect dense coding in the absence of noise and we refer to it as $|W_{1/2}\rangle^{3}$. For four-qubits, we can write it as $|gW\rangle^4 = \sqrt{a}|0001\rangle + \sqrt{b}|0010\rangle + \sqrt{c}|0100\rangle + \sqrt{1 - a - b - c}|1000\rangle$ (shared between $\mathcal{S}_1, \mathcal{S}_2, \mathcal{S}_3$ and $\mathcal{R}$) which has one ebit (entangled-bit - where one ebit refers to the amount of entanglement in a maximally entangled state of two parties \cite{Bennett_PRA_1996}) of entanglement for $a = 1/2$ in the $\mathcal{S} : \mathcal{R}$ bipartition  and we term it as $|W_{1/2}\rangle^4$.
For these classes of states, we
study the trends of  DCC against the noise strength and the variation of state parameters, for different values of the non-Markovianity parameter, \(\alpha\). In the case of the dephasing channel, the optimal unitary $U^{\min}$ is identified numerically (using the NLOPT algorithm ISRES \cite{NLOPT}).  \\

\textit{Two vs three senders with gGHZ states.}  When three- and four-qubit gGHZ states are shared between two and three senders respectively (with one receiver) and after encoding, the senders' qubits are disturbed by dephasing noise, we obtain the noisy encoded state, denoted by $\rho^{gGHZ}_{(p,\alpha)}$ with $\rho^{gGHZ}_{(0,0)}$ being the original pure state. As one expects, $C^1_{\text{noise}}(\rho^{gGHZ}_{(p,0)}) < C^1_{\text{noise}}(\rho^{gGHZ}_{(0,0)})$, i.e., the DCC decreases monotonically with the increase of the noise parameter $p$, irrespective of the state parameter. There exists a critical noise value $p_c$, where $C^1_{\text{noise}}(\rho^{gGHZ}_{(p,0)})$ reduces to its classical bound. For example, at $\alpha = 0$, $p_c = 0.48$ for the three-qubit gGHZ state while for four-qubit $|gGHZ\rangle^4$,  $p_c = 0.42$. Clearly, $p_c$ decreases with the number of senders, thereby showing a decrease in the robustness  against noise with an increasing number of parties. \\

\begin{figure}[!ht]
		\centering
		\includegraphics[width=\linewidth]{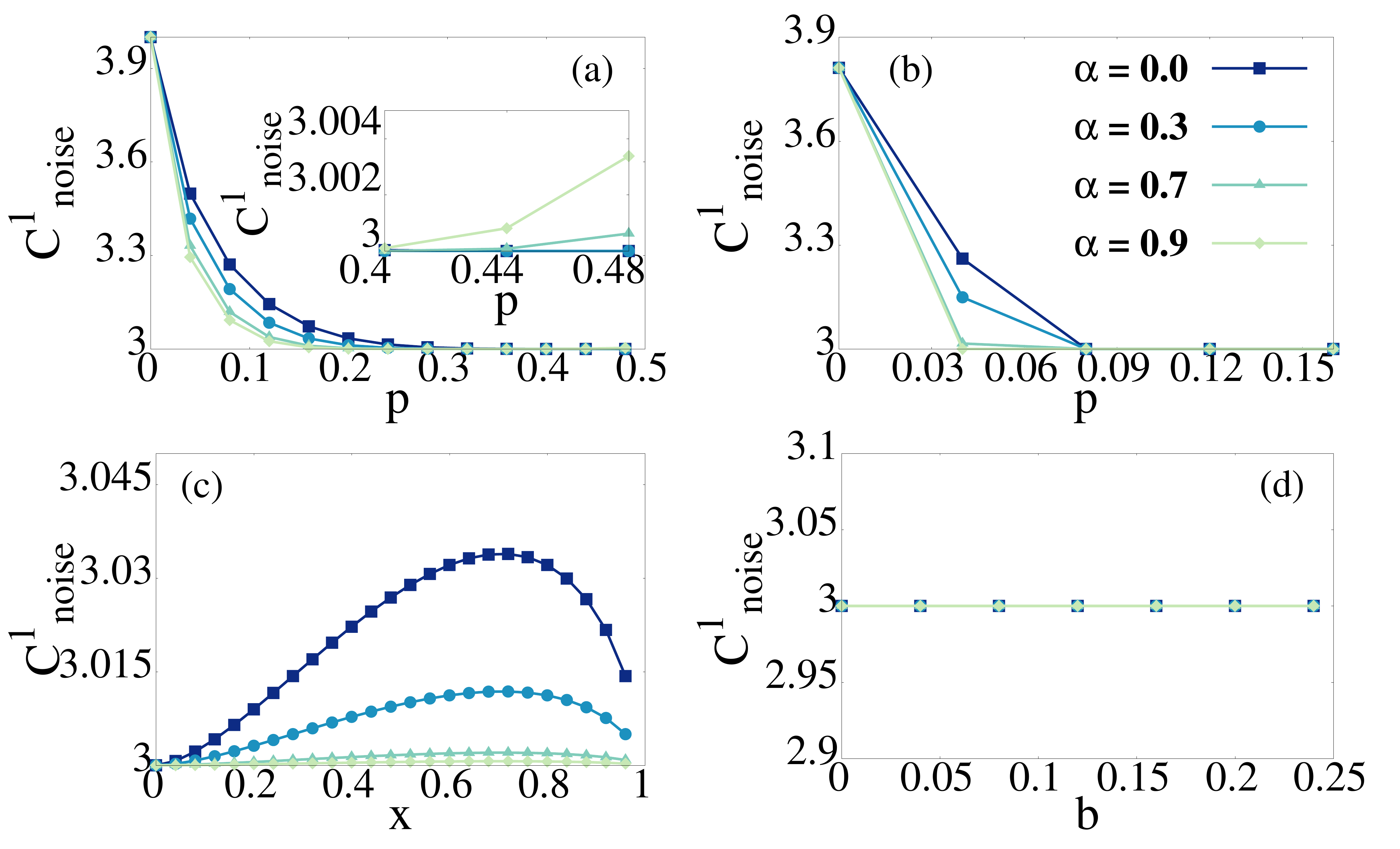}
		\caption{ (Color online.)  {\bf \(3 \mathcal{S} - 1\mathcal{R}\)  under non-Markovian dephasing noise.} (a) and (b). $C^1_{\text{noise}}$ (ordinate) vs $p$ (abscissa) when four-qubit GHZ and W states are initially shared. The constructive effect of non-Markovianity at high $p$ and $\alpha$ is highlighted in the inset in (a). DCC for the four-qubit $|gGHZ\rangle^{4}$ and $|W_{1/2}\rangle^{4}$ states with the state parameters, $x$ and $b$ (abscissa) respectively for $p = 0.4$ are shown in (c) and (d). All other specifications are the same as in Fig. \ref{fig:2s-1r-deph}.}
	\label{fig:3s-1r-deph}
\end{figure}

{\it Constructive effects of non-Markovianity.} Quantum advantage in the DC protocol emerges  with high noise in the presence of strong non-Markovianity, i.e., a higher non-Markovian strength allows for countering the destructive effects of noise on the DCC. Specifically, at high $p$ and $\alpha$, we observe the existence of \(p_a\), i.e., for $p \geq p_a$  $C^1_{\text{noise}}(\rho^{gGHZ}_{(p,\alpha)}) > C^1_{\text{noise}}(\rho^{gGHZ}_{(p,0)})$, thereby establishing the constructive response of non-Markovianity (see Figs. \ref{fig:2s-1r-deph} (a),  \ref{fig:2s-1r-deph} (c), and \ref{fig:3s-1r-deph} (a) and Table. \ref{table_nm_deph}). The advantage in DCC is more pronounced at high values of noise strength and non-Markovianity. In other words, $p_a$, as defined in Eq. \eqref{eq:pa}, decreases with $\alpha$ and \(p\). For the shared four-qubit gGHZ states, the quantum advantage can again be reported which are of two kinds - (i) $C^1_{\text{noise}}(|gGHZ\rangle^{4}_{(p,\alpha)})$ becomes non-vanishing after it collapses, i.e., $p_r$ exists like the three qubit case. (ii) Secondly, $C^1_{\text{noise}}(\rho^{gGHZ}_{(p,\alpha)}) > C^1_{\text{noise}}(\rho^{gGHZ}_{(p,0)})$, thereby obtaining $p_a$. Notice that, for the existence of $p_a$, a high amount of non-Markovianity is required for three senders compared to DC with two senders. It possibly indicates that when there are more senders, noise acts on each channel, thereby accumulating more destructive effects of noise which can only be eliminated with high non-Markovianity as shown in Table. \ref{table_nm_deph}. \\




\textit{No advantage of non-Markovianity for generalized W states.} In general, the DCC for the gW states, in the $N \mathcal{S}-1 \mathcal{R}$ routine, neither revives after collapse nor shows advantage due to non-Markovian noise after collapse (see Table. \ref{table_nm_deph}). 
In particular, as shown in Figs. \ref{fig:2s-1r-deph} (b), and \ref{fig:3s-1r-deph} (b) $p_c$ decreases with the strength of non-Markovianity $\alpha$, thereby illustrating less robustness of the $|W\rangle^3$ state against noise.  However, the $|W_{1/2}\rangle^{3}$ states show an opposite behavior, specifically a constructive effect with $\alpha$. For example, for $p = 0.4$ and sufficiently high non-Markovianity ($\alpha >0.8)$, we have $C^1_{\text{noise}}(|W_{1/2}\rangle^{3}_{(p,\alpha)}) > C^1_{\text{noise}}(|W_{1/2}\rangle^{3}_{(p,0)})$. In spite of such advantage, states for which non-classical capacity does not exist under Markovian noise ($b \geq 0.02$ as illustrated in Fig. \ref{fig:2s-1r-deph}(d) and for all $b$ in Fig. \ref{fig:3s-1r-deph} (d)), fail to exhibit quantum advantage at $\alpha >0$. \\

{\it Detrimental behavior observed for the non-Markovian depolarising channel.}
Like on entanglement,  the depolarising channel has a much more adverse effect on the dense coding protocol than the dephasing one which is clear from the very low value of \(p_c\) reported in  Table. \ref{table_nm_depo}. This is primarily due to the  fact that the depolarising noise acts on the qubits from all directions in contrast to the dephasing type noise which only affects the qubits from the z-direction \cite{Gupta_PRA_2022}. Neither the gGHZ nor the gW states show any advantage with $\alpha$, rather, their capacities decrease with non-Markovianity. As argued before, non-Markovianity can counter the destructive effects of noise only when the noise strength is above a certain threshold value which is not possible for the depolarizing channel since the noise parameter, $p$, is upper bounded by  $1/3\alpha$ for any arbitrary shared state.  Moreover, comparing the $2\mathcal{S}-1\mathcal{R}$ and the $3\mathcal{S}-1\mathcal{R}$ protocols, we find that the DCC is much worse affected for the three senders' case than that of the two senders' protocol (compare Figs. \ref{fig:3s-1r-deph} and \ref{fig:2s-2r-depo}). 

\subsection{Distributed non-Markovian dense coding}
\label{subsec:2-receiver_nm}
\begin{figure}[!ht]
		\centering
		\includegraphics[width=\linewidth]{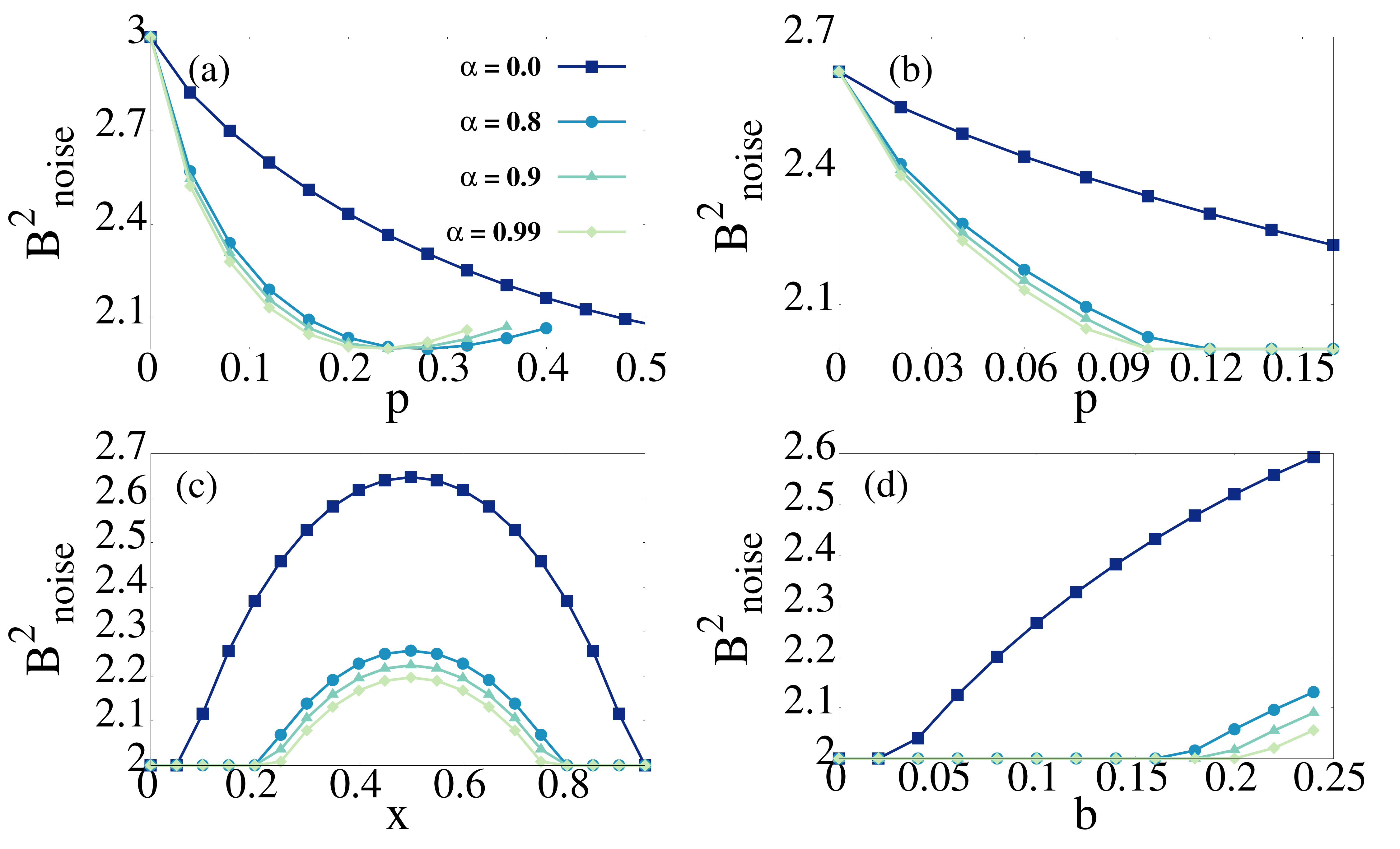}
		\caption{ (Color online.) {\bf Two senders and two receivers scenario.}  $B^{2}_{\text{noise}}$ with non-Markovian depolarising noise, \(p\) in (a) and (b) for the GHZ and W states while in (c) and (d), the upper bound on capacity is plotted against state parameters of $|gGHZ\rangle^{4}$ and \(|W_{1/2}\rangle^{4}\). 
		All other specifications are the same as in Fig. \ref{fig:2s-1r-deph}.}
	\label{fig:2s-2r-depo}
\end{figure}

A dense coding scheme with two senders and two receivers  is qualitatively different than the scenario with a single receiver. The impact  of noise on the distributed dense coding protocol can only be inferred by studying the pattern of an upper bound \(B^2_{noise}\) via LOCC  in Eq. (\ref{eq:DCC_2R_noisy}). 


We again consider the non-Markovian dephasing and depolarizing channels as noise models to obtain the expressions for the dense coding capacity given in Eq. \eqref{eq:DCC_2R_noisy}. In contrast to the single receiver scenario, we will show that the two-receiver dense coding protocol is more robust to the non-Markovian dephasing noise (since $p_c$ does not exist for the gGHZ states in the $2\mathcal{S}-2\mathcal{R}$ regime), and 
 the $|W_{1/2}\rangle^{4}$ states also provide a larger quantum advantage in DCC than the one-receiver scenario for both types of noise. Moreover, for the gGHZ states, the dephasing channel has no effect on the dense coding capacity when varied against $\alpha$ i.e., $B^2_{\text{noise}}(|gGHZ\rangle^4)$ is invariant for the entire range of non-Markovianity at a fixed noise strength $p$.\\
 
\textbf{Theorem 3}. \textit{For the four-qubit gGHZ state shared between two senders and two receivers, the  upper bound on the distributed DCC without noise coincides with the upper bound in the presence of non-Markovian dephasing noise, i.e., $B^2_{\text{noise}} = B^2$ irrespective of the value of the non-Markovian parameter $\alpha$}.\\

\textit{Proof}. The four-qubit gGHZ state, $|gGHZ\rangle^4_{\mathcal{S}_1 \mathcal{S}_2 \mathcal{R}_1 \mathcal{R}_2} = x|0000\rangle + \sqrt{1-x^2}|1111\rangle$ can be represented in the density matrix form as $\rho_{gGHZ} = x^2 |0000\rangle\langle0000| + (1-x^2)|1111\rangle\langle1111| + x\sqrt{(1-x^2)}(|0000\rangle\langle1111| + |1111\rangle\langle0000|)$. If we again assume that the $U^{\min}$ are proportional to the identity operator, which our numerical studies show to be the case, the non-Markovian noise does not change the subspace of the affected state, i.e., $\Lambda(\rho_{gGHZ}) = x^2 |0000\rangle\langle0000| + (1-x^2)|1111\rangle\langle1111| + f(p,\alpha,x)(|0000\rangle\langle1111| + |1111\rangle\langle0000|)$ where  the function $f$ includes the noise parameters $p$ and $\alpha$. Notice that only the off-diagonal terms change, which does not contribute during the partial trace operation, unlike the single receiver protocol. The only term in the expression of $B^2_{\text{noise}}$ that can get modified by the effect of noise is $\max [S(\text{Tr}_{\mathcal{S}_1 \mathcal{R}_1}\Lambda (\rho_{gGHZ})) , S(\text{Tr}_{\mathcal{S}_2 \mathcal{R}_2}\Lambda(\rho_{gGHZ}))]  = H(\{x^2, 1 - x^2, 0, 0\}) $ 
which is independent of the non-Markovianity parameter $\alpha$ and also the noise strength $p$. Thus independent of the value of $\alpha$ and $p$, the upper bound on the $2\mathcal{S}-2\mathcal{R}$ DCC remains unchanged and equal to its noiseless value $B^2_{\text{noise}} = 2 + H(\{x^2, 1 - x^2\})$ bits. Hence the proof. $\hfill \blacksquare$\\
\textbf{Remark}. The optimizing unitaries $U^{\min}$ appearing in Eq. \eqref{eq:DCC_2R_noisy} do not make any qualitative changes to our results since numerical studies suggest that each $U^{\min}$ acting on the sender's end, is proportional to the identity operator, and thus do not change the subspace to which the initial state belongs.\\
Let us now elaborate on our numerical observations pertaining to the noisy distributed dense coding scheme.
  \begin{itemize}
      \item \textbf{gGHZ states - dephasing vs depolarizing noise.} Although no non-Markovian advantage is reported for the dephasing channel in the entire parameter range of $\alpha$ for the gGHZ state, the GHZ state remains unaffected by non-Markovianity, furnishing $B^2_{\text{noise}} = 2 + H(\{x^2, 1 - x^2\})$ for all $\alpha$.  There is, however, a destructive effect on the DCC of the gGHZ states against depolarising noise, such that $B^2_{\text{noise}}$ decreases with $\alpha > 0$ although it remains above the classical threshold as illustrated in Figs. \ref{fig:2s-2r-depo}(a) and (c). 
      
      \item \textbf{gW states.} 
      For the dephasing type of noise, the dense coding capacity of the W state does not hit the classical value at all but decreases to a minimum value and the quantum advantage also increases with $\alpha$ as shown in Table. \ref{table_nm_deph}. On the contrary, the depolarising noise is again destructive in nature whence the capacity decreases with an increase of the non-Markovian parameter. The W state, according to Fig. \ref{fig:2s-2r-depo} (b) shows no advantage of non-Markovianity for the depolarising channel and the capacity collapses to its classical limit faster than in the Markovian regime. The $|W_{1/2}\rangle^{4}$ state also mimics this behavior, where the DCC decreases with $\alpha$ and the parameter regime of $b$ where quantum advantage exists also shrinks with non-Markovianity as shown in Fig. \ref{fig:2s-2r-depo}(d).
  \end{itemize}  
\begin{table}[]
	\caption{Critical noise strength $p_c$ for GHZ and W states affected by non-Markovian depolarising noise. } 
	\begin{tabular}{|c|lrrrrr|}
\hline
$\alpha$                  & \multicolumn{6}{c|}{$p_c$}                                                                                                                                                                                                                                                                                                  \\ \hline
                          & \multicolumn{1}{l|}{}                              & \multicolumn{1}{l|}{}                              & \multicolumn{1}{l|}{}                              & \multicolumn{1}{l|}{}                              & \multicolumn{1}{l|}{}                              & \multicolumn{1}{l|}{}                              \\ \hline
                          & \multicolumn{3}{c|}{GHZ}                                                                                                                                     & \multicolumn{3}{c|}{W}                                                                                                                                       \\ \hline
                          & \multicolumn{1}{l|}{}                              & \multicolumn{1}{l|}{}                              & \multicolumn{1}{l|}{}                              & \multicolumn{1}{l|}{}                              & \multicolumn{1}{l|}{}                              & \multicolumn{1}{l|}{}                              \\ \hline
                          & \multicolumn{1}{c|}{$2 \mathcal{S}-1 \mathcal{R}$} & \multicolumn{1}{c|}{$3 \mathcal{S}-1 \mathcal{R}$} & \multicolumn{1}{c|}{$2 \mathcal{S}-2 \mathcal{R}$} & \multicolumn{1}{c|}{$2 \mathcal{S}-1 \mathcal{R}$} & \multicolumn{1}{c|}{$3 \mathcal{S}-1 \mathcal{R}$} & \multicolumn{1}{c|}{$2 \mathcal{S}-2 \mathcal{R}$} \\ \hline
                          & \multicolumn{1}{l|}{}                              & \multicolumn{1}{l|}{}                              & \multicolumn{1}{l|}{}                              & \multicolumn{1}{l|}{}                              & \multicolumn{1}{l|}{}                              & \multicolumn{1}{l|}{}                              \\ \hline
\multicolumn{1}{|r|}{0}   & \multicolumn{1}{r|}{0.09}                          & \multicolumn{1}{r|}{0.06}                          & \multicolumn{1}{r|}{0.75}                          & \multicolumn{1}{r|}{0.08}                          & \multicolumn{1}{r|}{0.05}                          & 0.31                                               \\ \hline
\multicolumn{1}{|r|}{0.3} & \multicolumn{1}{r|}{0.07}                          & \multicolumn{1}{r|}{0.03}                          & \multicolumn{1}{r|}{0.58}                          & \multicolumn{1}{r|}{0.05}                          & \multicolumn{1}{r|}{0.03}                          & 0.26                                               \\ \hline
\multicolumn{1}{|r|}{0.5} & \multicolumn{1}{r|}{0.05}                          & \multicolumn{1}{r|}{0.03}                          & \multicolumn{1}{r|}{0.45}                          & \multicolumn{1}{r|}{0.04}                          & \multicolumn{1}{r|}{0.02}                          & 0.21                                               \\ \hline
\multicolumn{1}{|r|}{0.7} & \multicolumn{1}{r|}{0.04}                          & \multicolumn{1}{r|}{0.02}                          & \multicolumn{1}{r|}{0.32}                          & \multicolumn{1}{r|}{0.03}                          & \multicolumn{1}{r|}{0.02}                          & 0.16                                               \\ \hline
\multicolumn{1}{|r|}{0.9} & \multicolumn{1}{r|}{0.03}                          & \multicolumn{1}{r|}{0.02}                          & \multicolumn{1}{r|}{0.25}                          & \multicolumn{1}{r|}{0.02}                          & \multicolumn{1}{r|}{0.02}                          & 0.1                                                \\ \hline
\end{tabular}
	\label{table_nm_depo}
\end{table}
\section{Effects of random channels on Dense Coding Capacity}
\label{sec:random_channels}

In this section, we study the impact of noise on the dense coding protocol, when the noisy channels are characterized by random unitaries along with non-Markovianity as discussed in Sec. \ref{subsec:random_noise}. Note that in the case of random noise, the DCC has to be computed by performing averaging which we define now.

{\it Quenched averaging.} For a fixed resource state and non-Markovianity $\alpha$, in a single realization, we compute $C^{1}_{noise}$ or $B^{2}_{noise}$ by choosing a set of parameters in the unitary, $\{x_{i}\}$, from a Gaussian distribution with mean $\langle x_{i}\rangle$ and standard deviation $\epsilon$.   We calculate the dense coding capacity for $4 \times 10^{3}$ sets of such realizations and by averaging them,  we obtain quenched averaged dense coding capacity denoted by $\langle C^{1}_{noise} \rangle$ or  $\langle B^{2}_{noise} \rangle$. \\

\textbf{Theorem 4.} \textit{The upper bound on the dense coding capacity affected by the random noisy channel for multiple senders and a single receiver  is greater than the capacity influenced by  noise without randomness.}\\

\textit{Proof}. Any arbitrary unitary operator in two dimensions can be characterized by three parameters $\omega, \theta$, and $\delta$, apart from an overall phase, as in Eq. \eqref{eq:random_unitary}. Let us suppose that any Pauli matrix is written as $U^o = U_1^o U_2^o U_3^o$, characterized by the parameters $\omega^o, \theta^o$, and $\delta^o$ respectively. Suppose, the random unitary corresponding to $U^o$ is $U = U_1 U_2 U_3$. To generate such a random unitary, each parameter, ($\omega, \theta$, and $\delta$) is randomly chosen from a Gaussian distribution of mean $\omega^o, \theta^o$, and $\delta^o$ respectively and standard deviation $\epsilon$.
Thus one arbitrary choice of the parameters of the random unitary $U$ may be taken to be $\omega^o + \epsilon, \theta^o + \epsilon, \delta^o + \epsilon$. It can be shown that the unitaries constituting $U$ can be written as $U_i = \cos \frac{\epsilon}{2} U_i^o + \sin \frac{\epsilon}{2} U_i'$ with $i = 1, 2, 3$ where 
\begin{eqnarray}
   && U_1' = \begin{pmatrix}
    i e^{i\omega/2} & 0 \\
    0 & -i e^{-i\omega/2}
    \end{pmatrix}, \nonumber \\
    && U_2' = \begin{pmatrix}
    -\sin \theta/2 & \cos \theta/2 \\
    -\cos \theta/2 & - \sin \theta/2
    \end{pmatrix}, \label{eq:proof_random1b} \nonumber \\
    && U_3' = \begin{pmatrix}
    i e^{i\delta/2} & 0 \\
    0 & -i e^{-i\delta/2}
    \end{pmatrix}. \label{eq:proof_random1c}
\end{eqnarray}
Thus  $U = U_1 U_2 U_3 =  \cos^3 \frac{\epsilon}{2} U^0 + \sum_j f_j (\epsilon) \tilde{U}_j$, where each $f_j(\epsilon) = \mathcal{O}(\sin \frac{\epsilon}{2})$ or higher and each $\tilde{U}_j$ represents a unitary obtained from the product of $U_i's$ and $U^{o}_{i}$'s. Note that we denote $\tilde{U}_0 = U_1^0 U_2^0 U_3^0 = U^0$ which is not present in the summation and thus $j > 0$. Finally, we can write the action of $U$ on a state $\rho$ as
\begin{eqnarray}
   \nonumber   U \rho U^\dagger && = \cos^6 \frac{\epsilon}{2} U^o \rho U^{o \dagger} + \sum_{j>0} f_j(\epsilon) U^o \rho \tilde{U}_j^\dagger + \sum_{j>0} f_j(\epsilon) \tilde{U}_j \rho U^{o \dagger} \\
   && + \sum_{j>0,l>0} f_j(\epsilon) f_l(\epsilon) \tilde{U}_j \rho \tilde{U}_l^\dagger  \label{eq:proof_random2a}\\
   && = \cos^6 \frac{\epsilon}{2} U^o \rho U^{o \dagger} + \sum_{j,l} f_j (\epsilon) f_l(\epsilon) \tilde{U}_j \rho \tilde{U}_l^\dagger. \label{eq:proof_random2b}
    \end{eqnarray}
Here, one must  take care to omit $j = l = 0$ in the summation.\\

Let us now illustrate the proof for the dephasing channel. Note that  the same line of proof would hold for the depolarising channel, but with some more cumbersome algebra. The dephasing channel is characterized by the Pauli operator $U^o = \sigma_z$ whose parameters $\omega^o, \theta^o$ and $\delta^o$ are given in Sec. \ref{subsec:random_noise}. Since we consider the noise to act only on the senders' part, the only affected term in the dense coding capacity given by Eq. \eqref{eq:DCC_3-1_noisy} is $S(\tilde{\rho})$. In the presence of a random dephasing channel, we can write
 \(   S(\tilde{\rho}_{random}) = S(p_1 \rho + p_2 U \rho U^\dagger)\),
where $p_1 = (1 - \alpha p)(1-p)$ and $p_2 = (1 + \alpha (1 - p))p$ with $p$ being the noise parameter of the dephasing channel and $\alpha$, the strength of non-Markovianity. Using  concavity of entropy, \cite{nielsenchuang}, i.e.,
 \(S(\sum_i p_i \rho_i) \leq \sum_i p_i S(\rho_i) + H\{p_i\}\),
we obtain
\begin{eqnarray}
   && S(p_1 \rho + \cos^6 \frac{\epsilon}{2} p_2 U^o \rho U^{o \dagger} + p_2 \sum_{\substack{i,j\\j \neq i}} f_i (\epsilon) f_j(\epsilon) U_i \rho U_j^\dagger) \\
    && \nonumber  \leq S(p_1 \rho + \cos^6 \frac{\epsilon}{2} p_2 U^o \rho U^{o \dagger}) + p_2 \sum_{\substack{i,j\\j \neq i}} f_i (\epsilon) f_j(\epsilon) S(U_i \rho U_j^\dagger) \\
    && ~~~~~~~~~~~~~~~~~~~~~~~~ + H\{p_1, p_2, p_2 f_i(\epsilon) f_j (\epsilon)\} .  \label{eq:proof_random5}
    \end{eqnarray}
    We can neglect the last summation, since each $f_i(\epsilon)$ is a function of $\sin \epsilon$ and we have assumed $\epsilon \to 0$, which also implies $\cos^6 \epsilon \to 1$. Thus we finally arrive at 
    \begin{eqnarray}
       && S(\tilde{\rho}_{random}) \leq S(\tilde{\rho}_{dph}) + H\{p_1, p_2\} \\
      && \nonumber  \implies C^1_{\text{noise}}(\Lambda_{\text{random}}(\rho)) \geq C^1_{\text{noise}}(\Lambda_{\text{dph}}(\rho)) - \\
     &&~~~~~~~~~~~~~~~~~~~~~~~~~~~~~~~~~~~~~~~~~~~~~~~~~ H\{p_1, p_2\}.
        \label{eq:proof_random6}
    \end{eqnarray}
Numerical simulations confirm that $S(\tilde{\rho}_{random}) \leq S(\tilde{\rho}_{dph})$ for small as well as moderate disorder strength, \(\epsilon\). 
Therefore, the dense coding capacity for a state affected by random noise is greater than that affected by Pauli noise.  $\hfill \blacksquare$

\textbf{Remark.} In case of dephasing channel,  
$U^{\min}$ is again chosen to be the identity operator in the proof for simplicity. However,
numerical simulations again show that the results remain true even when optimization over unitaries are performed.  

\begin{table*}[]
	\caption{Critical noise strength for collapse, $p_c$, for GHZ and W states affected by random depolarising noise. } 
	\begin{tabular}{|c|cccccc|cccccc|}
\hline
                          & \multicolumn{6}{c|}{GHZ}                                                                                                                                                                                                          & \multicolumn{6}{c|}{W}                                                                                                                                                                                                            \\ \hline
                          & \multicolumn{1}{c|}{}                 & \multicolumn{1}{c|}{}                 & \multicolumn{1}{c|}{}                 & \multicolumn{1}{c|}{}                 & \multicolumn{1}{c|}{}                 &                           & \multicolumn{1}{c|}{}                 & \multicolumn{1}{c|}{}                 & \multicolumn{1}{c|}{}                 & \multicolumn{1}{c|}{}                 & \multicolumn{1}{c|}{}                 &                           \\ \hline
                          & \multicolumn{3}{c|}{$2 \mathcal{S}-1 \mathcal{R}$}                                                                    & \multicolumn{3}{c|}{$3 \mathcal{S}-1 \mathcal{R}$}                                                        & \multicolumn{3}{c|}{$2 \mathcal{S}-1 \mathcal{R}$}                                                                    & \multicolumn{3}{c|}{$3 \mathcal{S}-1 \mathcal{R}$}                                                        \\ \hline
                          & \multicolumn{1}{c|}{}                 & \multicolumn{1}{c|}{}                 & \multicolumn{1}{c|}{}                 & \multicolumn{1}{c|}{}                 & \multicolumn{1}{c|}{}                 &                           & \multicolumn{1}{c|}{}                 & \multicolumn{1}{c|}{}                 & \multicolumn{1}{c|}{}                 & \multicolumn{1}{c|}{}                 & \multicolumn{1}{c|}{}                 &                           \\ \hline
$\alpha$                  & \multicolumn{1}{c|}{$\epsilon = 0.5$} & \multicolumn{1}{c|}{$\epsilon = 0.7$} & \multicolumn{1}{c|}{$\epsilon = 1.0$} & \multicolumn{1}{c|}{$\epsilon = 0.5$} & \multicolumn{1}{c|}{$\epsilon = 0.7$} & $\epsilon = 1.0$          & \multicolumn{1}{c|}{$\epsilon = 0.5$} & \multicolumn{1}{c|}{$\epsilon = 0.7$} & \multicolumn{1}{c|}{$\epsilon = 1.0$} & \multicolumn{1}{c|}{$\epsilon = 0.5$} & \multicolumn{1}{c|}{$\epsilon = 0.7$} & $\epsilon = 1.0$          \\ \hline
\multicolumn{1}{|l|}{}    & \multicolumn{1}{l|}{}                 & \multicolumn{1}{l|}{}                 & \multicolumn{1}{l|}{}                 & \multicolumn{1}{l|}{}                 & \multicolumn{1}{l|}{}                 & \multicolumn{1}{l|}{}     & \multicolumn{1}{l|}{}                 & \multicolumn{1}{l|}{}                 & \multicolumn{1}{l|}{}                 & \multicolumn{1}{l|}{}                 & \multicolumn{1}{l|}{}                 & \multicolumn{1}{l|}{}     \\ \hline
\multicolumn{1}{|r|}{0.3} & \multicolumn{1}{r|}{0.09}             & \multicolumn{1}{r|}{0.11}             & \multicolumn{1}{r|}{0.14}             & \multicolumn{1}{r|}{0.04}             & \multicolumn{1}{r|}{0.05}             & \multicolumn{1}{r|}{0.06} & \multicolumn{1}{r|}{0.08}             & \multicolumn{1}{r|}{0.09}             & \multicolumn{1}{r|}{0.12}             & \multicolumn{1}{r|}{0.03}             & \multicolumn{1}{r|}{0.04}             & \multicolumn{1}{r|}{0.05} \\ \hline
\multicolumn{1}{|r|}{0.5} & \multicolumn{1}{r|}{0.06}             & \multicolumn{1}{r|}{0.08}             & \multicolumn{1}{r|}{0.1}              & \multicolumn{1}{r|}{0.03}             & \multicolumn{1}{r|}{0.04}             & \multicolumn{1}{r|}{0.05} & \multicolumn{1}{r|}{0.05}             & \multicolumn{1}{r|}{0.06}             & \multicolumn{1}{r|}{0.09}             & \multicolumn{1}{r|}{0.03}             & \multicolumn{1}{r|}{0.03}             & \multicolumn{1}{r|}{0.04} \\ \hline
\multicolumn{1}{|r|}{0.9} & \multicolumn{1}{r|}{0.04}             & \multicolumn{1}{r|}{0.05}             & \multicolumn{1}{r|}{0.07}             & \multicolumn{1}{r|}{0.02}             & \multicolumn{1}{r|}{0.03}             & \multicolumn{1}{r|}{0.04} & \multicolumn{1}{r|}{0.03}             & \multicolumn{1}{r|}{0.04}             & \multicolumn{1}{r|}{0.07}             & \multicolumn{1}{r|}{0.02}             & \multicolumn{1}{r|}{0.02}             & \multicolumn{1}{r|}{0.01} \\ \hline
\end{tabular}
	\label{tab:p_c}
\end{table*}
\subsection{Random dephasing noise on DCC}
\label{subsec:rand_deph}
To begin our discussion on the effect of random noisy channels, we first consider the dephasing channel composed of two Kraus operators, one involving the identity, and the other involving $\sigma_{z}$. In the case of random channels, the parameters of unitaries are chosen from a Gaussian distribution of standard deviation $\epsilon$ and mean fixed to the parameters of the corresponding Pauli matrix. We notice that the presence of such random noise provides a distinct advantage in the dense coding protocol in comparison with the effects of a pure dephasing channel, i.e., $C^{1}_{\text{noise}}(\Lambda_{p,\alpha,\epsilon}(\rho)) \geq C^{1}_{\text{noise}}(\Lambda_{p,\alpha,0}(\rho))$, where $\epsilon$ denotes the magnitude of randomness present in the noise. Let us present our observations for the case of single- and two receivers scenarios.  
\begin{figure}[!ht]
		\centering
		\includegraphics[width=\linewidth]{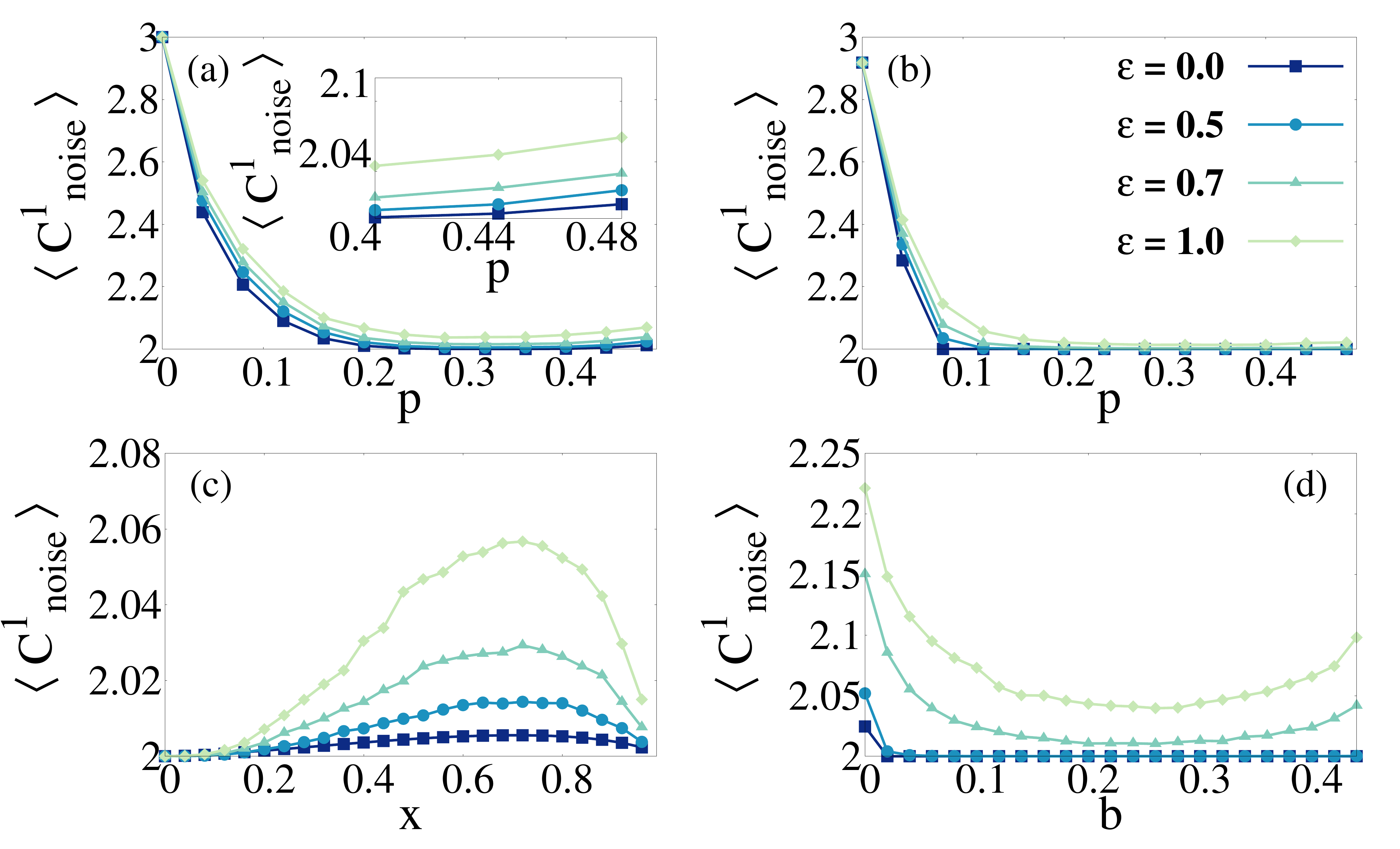}
		\caption{ (Color online.) {\bf \(2 \mathcal{S}-1 \mathcal{R}\) DC with random non-Markovian dephasing noise.  } Quenched averaged dense coding capacity, $\langle C^{1}_{\text{noise}}\rangle$, (ordinate) with different disorder strength, $\epsilon$  against the noise parameter $p$ in (a). and (b)., and against the state parameter $x$ and $b$ (abscissa) in (c). and (d) respectively. The states shared between senders and the receiver are the three-qubit GHZ (in (a)), W (in (b)), $|gGHZ\rangle^{3}$ (in (c)), and $|W_{1/2}\rangle^{3}$ states (in (d)).  Squares, circles, triangles and diamonds represent $\epsilon = 0$,  \(\epsilon =0.5\),  $\epsilon = 0.7$ and $\epsilon = 1.0$   respectively. In all the cases, the non-Markovianity parameter is fixed to $\alpha = 0.8$. The horizontal axis is dimensionless although the vertical axis is in bits.}
	\label{fig:random_1r_deph}
\end{figure}

\subsubsection{Random channel-effected  dense coding involving a single receiver: Comparison between gGHZ and gW states}
\label{subsubsec:random_deph_1r}

\begin{figure*}[!ht]
		\centering
		\includegraphics[width=\linewidth]{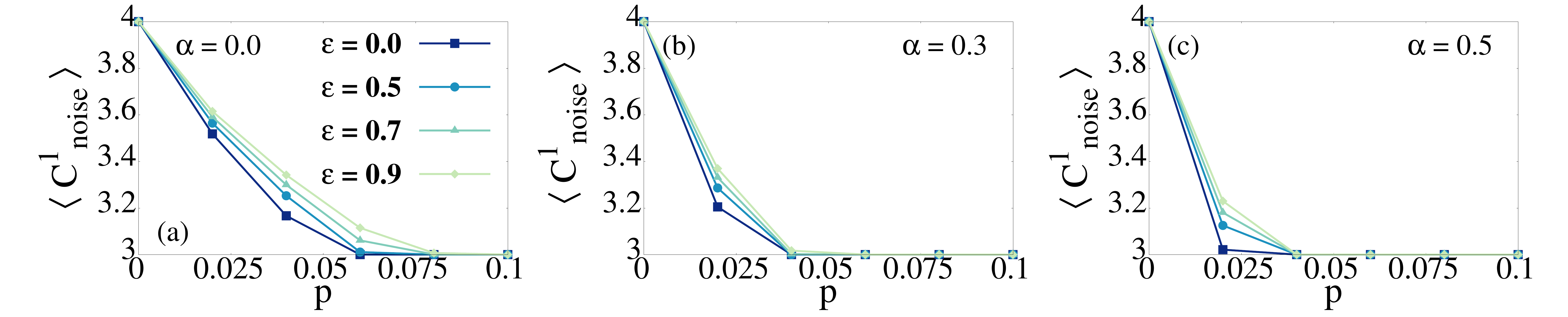}
		\caption{(Color online.) 
		{\bf Advantage in random depolarizing noise in \(3\mathcal{S}- 1\mathcal{R}\) DC protocol.}
		\(\langle C^1_{\text{noise}}\rangle \) (ordinate)
		against $p$ (abscissa) for different disorder strengths with the shared four-qubit GHZ state. 
		  Markovian noise, i.e., $\alpha = 0.0$,  (in (a)) is compared with non-Markovian noise, $\alpha = 0.3$ (in (b)) and $\alpha = 0.5$ (in (c)) respectively. All other specifications are the same as in Fig. \ref{fig:random_1r_deph}.}
	\label{fig:random_1r_depo}
\end{figure*}

\begin{figure}[!ht]
		\centering
		\includegraphics[width=\linewidth]{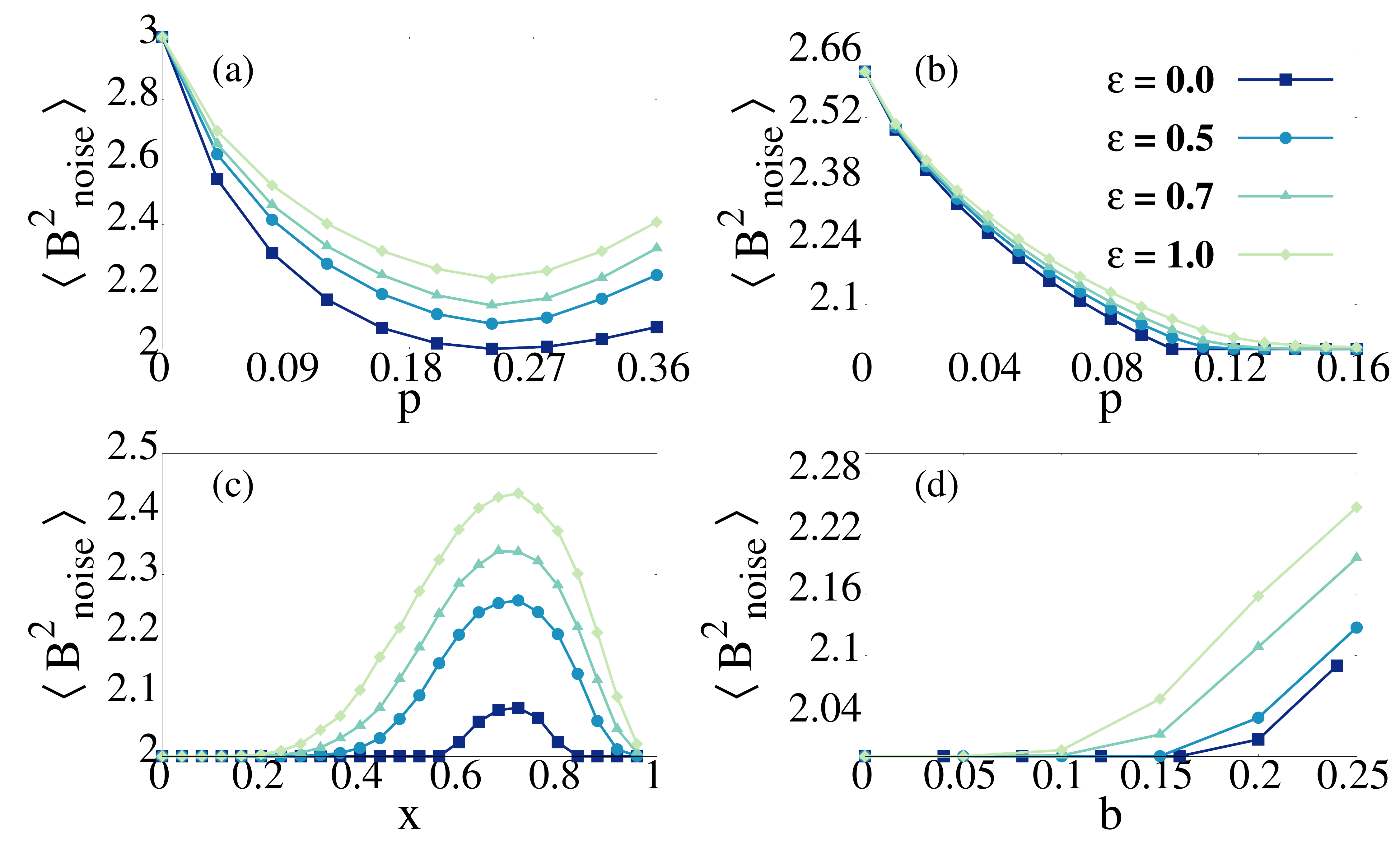}
		\caption{ (Color online.) The average upper bound on DCC  under random non-Markovian depolarising noise when $\alpha = 0.9$. All other specifications are the same as in Fig. \ref{fig:random_1r_deph}.}
	\label{fig:random_2r_depo}
\end{figure}

 \noindent \textbf{$2\mathcal{S}$-$1\mathcal{R}$ case}. In the case of the three-qubit GHZ state, we find that the average capacity increases steadily with $\epsilon$ for any value of $\alpha$ as shown in Fig. \ref{fig:random_1r_deph} (a). In fact, for sufficiently high standard deviation, e.g. $\epsilon \geq 0.7$, the quenched average dense coding capacity does not collapse to its classical threshold (which is $2$) but reaches a minimum value ($> 2$) and then rises again with the noise strength $p$. In this case, $p_a$ decreases with respect to both $\alpha$ and $\epsilon$. The decrease in $p_a$ with $\epsilon$  predicts the beneficial impact of randomness present in the noise acting on the protocol. Qualitatively similar behavior is observed when the W state is considered as the resource (see Fig. \ref{fig:random_1r_deph} (b).).
 This demonstrates that if the dephasing channel is not perfect, the protocol is much more beneficial in terms of dense coding capacity. 
 
 When analyzed with respect to the state parameter, we observe that the protocol becomes more and more efficient for a given $\alpha$ as the standard deviation $\epsilon$ assumes higher and higher values for the $|gGHZ\rangle^{3}$ and $|W_{1/2}\rangle^{3}$ states. For any given non-Markovianity strength and standard deviation, the capacity becomes maximum at $x = 1/\sqrt{2}$ (the GHZ state) as demonstrated in Fig. \ref{fig:random_1r_deph} (c). An important feature of randomly generated dephasing type noise emerges - there exist some $p$ and $x$ for which the Markovian pure dephasing channel offers no quantum advantage with the $|W_{1/2}\rangle^{3}$ state although an increase in the standard deviation makes the capacity overcome the classical limit. Thus the constructive effect of randomness is again highlighted in the noisy DC protocol.\\
 
\textbf{$3\mathcal{S}$-$1\mathcal{R}$ scenario}. The advantageous effect of randomness still persists when we consider $|GHZ\rangle^4$ but is much less pronounced (see Fig. \ref{fig:random_1r_depo}), with the revival of capacity beyond its classical value (upon collapse) being absent. Thus, with an increase in the number of senders, the destructive effect of the noise cannot be avoided through its random implementation, since more qubits are affected by it. In the protocol involving three senders sharing a W state with a lone receiver, no quantum advantage beyond that for the pure dephasing channel is apparent no matter how high the value of $\epsilon$ is. For the $|gGHZ\rangle^4$ states, the random noise cannot help in overcoming the classical limit of $3$ for any value of $\epsilon$ or $\alpha$. On the other hand, for sufficiently high randomness ($\epsilon \geq 0.5$), some quantum advantage is observed for $|W_{1/2}\rangle^{4}$.
\label{subsubsec:random_deph_2r}

\subsubsection{Effects of random noise on distributed dense coding}
\noindent When the dense coding protocol involves two receivers, we consider four qubit resource states shared between two senders and two receivers. The principal difference in the dense coding protocol involving two receivers, from that having a single receiver, is that the random noisy channel exhibits no effect when considering the gGHZ states. The constructive effect of randomness is observed for the W and $|W_{1/2}\rangle^{4}$ states, in Figs. \ref{fig:random_2r_depo} (b), (d), \ref{fig:random_2r_deph} (b) and (d) respectively, where the quenched averaged capacity increases with $\epsilon$ (however, it still remains lower than the gGHZ states). Unlike the gGHZ state, for which the region of quantum advantage is stretched across the entire parameter regime, the $|W_{1/2}\rangle^{4}$ states demonstrate quantum advantage only beyond $b \approx 0.1$.

\subsection{Depolarising random noise on dense coding: Beneficial role}
\label{subsec:depo_random}
We have already observed that depolarizing channels acting on the encoded qubits has extremely damaging effects on the performance of DC. It will be interesting to find out whether random channels can have a more destructive impact or less. Our results can be listed  according to the resource states.

\begin{enumerate}
    \item \textbf{gGHZ state as resource.} Although the pure Markovian depolarising channel suppresses any quantum advantage, an increase in $\epsilon$ at $\alpha = 0$ causes an increment in the quenched averaged capacity beyond its classical value both in the $2\mathcal{S}-1\mathcal{R}$ and $3\mathcal{S}-1\mathcal{R}$ regimes. The advantage in the dense coding capacity is pronounced around $x = 1/\sqrt{2}$ with the region of advantage increasing with an increase in the standard deviation (see Table. \ref{tab:p_c} for the GHZ state ($x = \frac{1}{\sqrt{2}}$)). However, the $\langle C^{1}_{noise}\rangle$ decreases with $\alpha$ for a given randomness parameter $\epsilon$ as demonstrated in Fig. \ref{fig:random_1r_depo} and Fig. \ref{fig:random_2r_depo} (c). Moreover, we observe that the critical noise strength of collapse, $p_c$, increases with the randomness, thereby sustaining non-classical capacity for a large range of the noise parameter.

The situation is similar when the four-qubit gGHZ state is shared in the $2\mathcal{S} - 2\mathcal{R}$ case. The averaged capacity here undergoes improvements with increasing $\epsilon$, as shown in Fig. \ref{fig:random_2r_depo} (a). The most important constructive feature is - for sufficient randomness in the channel ($\epsilon > 0.5$), the quenched average capacity never reduces to its classical value of $2$, but rises after reaching a minimum value. This is in contrast to the depolarising noise without randomness, where revival is observed after the capacity becomes $2$. When the capacity attains a minimum value, an increase in the standard deviation increases the noise strength at which the minimum occurs, thereby again exhibiting a constructive effect since the capacity remains high through a large noise parameter regime. 
\begin{figure}[!ht]
		\centering
		\includegraphics[width=\linewidth]{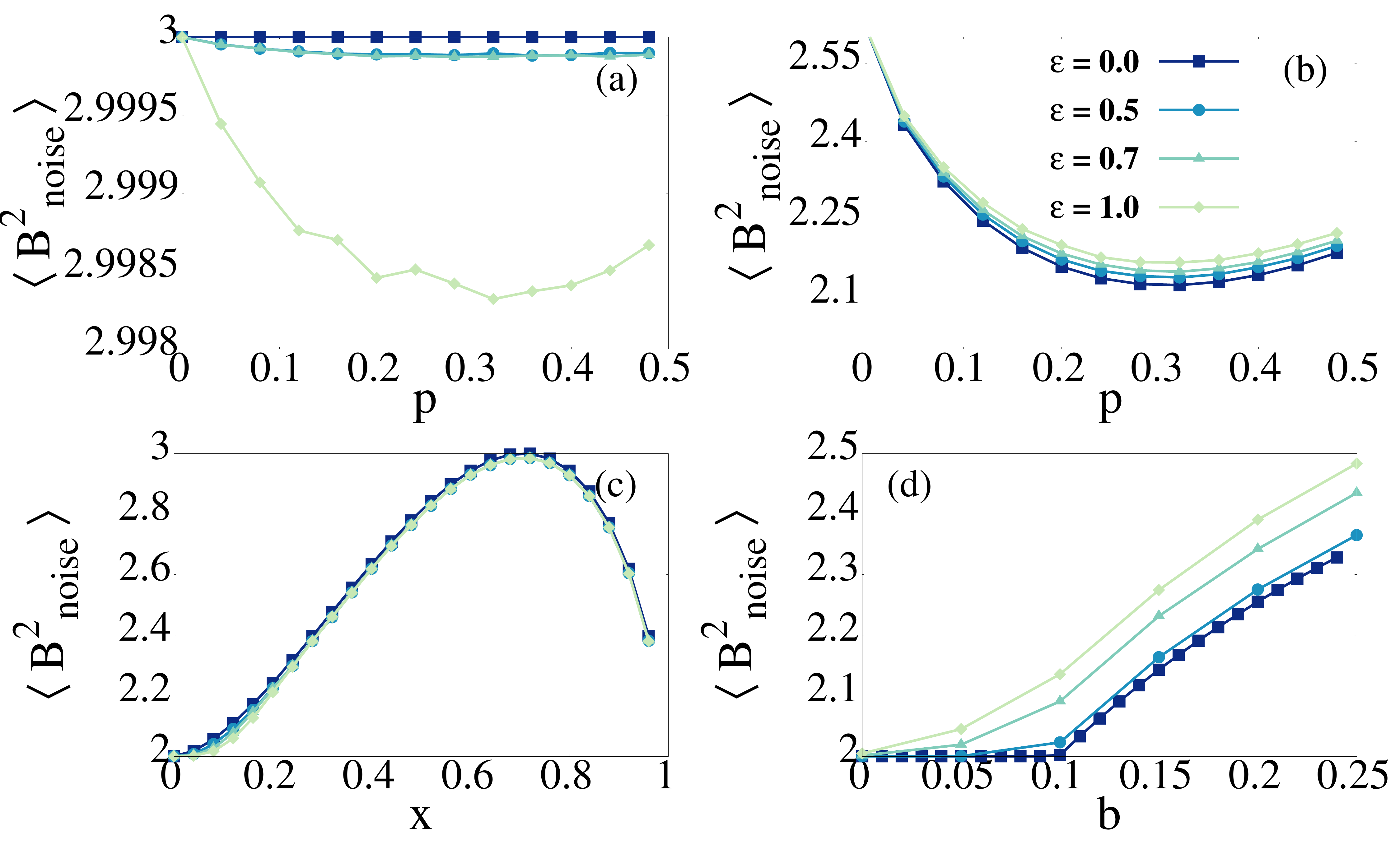}
		\caption{ (Color online.) The quenched averaged upper bound on the dense coding capacity, \(\langle B^2_{\text{noise}}\rangle\), (ordinate)  is demonstrated under random non-Markovian dephasing noise with $\alpha = 0.9$ in two senders-two receivers picture. All other specifications are the same as in Fig. \ref{fig:random_1r_deph}.}
	\label{fig:random_2r_deph}
\end{figure}
\item \textbf{gW state.} When a single receiver is involved in the dense coding protocol, the behavior of the averaged capacity of the gW state is similar to that of the gGHZ state, even though the quantum advantage is markedly less. The random noise allows for constructive effect in the two receivers scenario (see Fig. \ref{fig:random_2r_depo} (b).) whereas the pure depolarising channel always caused the capacity to decrease with $\alpha$, i.e., the random channel causes an increase in the DCC with increasing $\epsilon$.  Moreover, the probability of collapse $p_c$, also increases with the standard deviation (see Table. \ref{tab:p_c}). However, the noise strength of collapse is lower for the W state as compared to that for the GHZ state, thereby highlighting the increased susceptibility of the W state to noise. 

    \item \textbf{A special class of gW state.} For $|W_{1/2}\rangle^{3}$ and $|W_{1/2}\rangle^{4}$ shared with a single receiver, the dense coding capacity can never overcome its classical threshold in presence of non-Markovian noise. However, as illustrated in Fig. \ref{fig:random_2r_depo} (d), a constructive impact of randomness is apparent in the $2\mathcal{S}-2\mathcal{R}$ regime where the quenched average upper bound $\langle B^{2}_{noise}\rangle$ increases beyond the classical bound with an increase in $\epsilon$. Further, the state parameter region beyond which quantum advantage is obtained also expands with $\epsilon$ (for example at $\epsilon = 0.3, \langle B^{2}_{noise}\rangle > 2$ for $b \geq 0.15$ whereas $\langle B^{2}_{noise}\rangle > 2$ beyond $b = 0.1$ when $\epsilon = 0.7$), thereby providing a two-fold advantage in the protocol.
\end{enumerate}

\section{Conclusion}
\label{sec:conclusion}

Quantum superdense coding illustrates quantum advantage for transferring classical information encoded in quantum states, provided an entangled state is apriori shared between the senders and the receivers. \textcolor{black}{It is one of the first information theoretic protocols to be proposed and then experimentally realized \cite{Mattle_PRL_1996, Fang_PRA_2000, Wei_CSB_2004, Schaetz_PRL_2004}. During its realization, it is imperative that environmental impacts would induce unwanted noise in the system, thereby diminishing the performance. } To date, all the studies performed on the dense coding protocol have been solely focused on how Markovian noise affects the system's encoded component.  In a multipartite domain with arbitrary senders and one or two receivers, we found that when non-Markovian dephasing noise acts on the encoded part (especially for the three-party generalized Greenberger-Horne-Zeilinger(gGHZ) shared state), the dense coding capacity increases with an increase in non-Markovianity.  This non-Markovian enhancement is less prominent either when the shared state is the W-state or when one increases the number of parties.

\textcolor{black}{Noise is typically characterized by paradigmatic models such as the dephasing and depolarizing channels \cite{Preskill}.} From the experimental point of view, the actions of exact dephasing or depolarising noise involving Pauli matrices on the sender's part is an extremely idealized situation. \textcolor{black}{In reality, there must be some deviation from the Pauli noisy channel. Therefore, our work aimed to analyze how exactly the performance of the dense coding scheme would be altered when the noise itself behaved randomly.  This formalism introduces unknown stochastic elements into the implementation of the protocol, which are completely beyond the users' control.}  We addressed this question of how the dense coding capacity gets affected if we choose  Kraus representations involving unitaries from a Gaussian distribution with a mean around Pauli noise and a finite standard deviation. \textcolor{black}{Such a noise model, if present, would have a profound impact on the realization of the dense coding protocol, for example, it would destroy the covariant nature of the depolarising noise channel, thereby rendering the experiment invalid if the assumption of that particular channel is taken into account in the experiment. Furthermore, not knowing exactly which form of noise acts on the system would prevent the application of the entropy-minimizing unitaries which are crucial for obtaining quantum advantage. Therefore, it is intriguing to investigate the impact of random noise on the capacity of classical information transmission in networks, which is one of the important directions of research in the quantum communication field.} We dealt with multiple senders and single or two receivers sharing multipartite entangled states as the resource.  Most strikingly, we showed both analytically and numerically that in the presence of random noisy channels, the dense coding capacity increases with the increase of the strength of the randomness in the channel. \textcolor{black}{Our results showed that as we deviated from the Pauli nature of the noise, the protocol furnished better results in the sense that one can overcome the detrimental effects of noise on the protocol. However, on increasing the standard deviation of the Gaussian probability distribution, from which we sample the noise parameters, the noise would eventually revert to its Pauli characteristics at a certain point. Therefore, we investigated the interplay between the randomness present in the noise and the enhanced quantum advantage due to noise. In particular, quantum advantage actually persists if and only if the stochastic nature of the noise is within close range of the Pauli matrices. Therefore, we limited our analysis to standard deviations up to unity, which also hints at how one could construct the minimizing unitaries (by guessing the standard deviation, it is possible to incorporate its effects on the minimizing unitaries through simple numerical calculations) in the experiment. However, if the randomness in the noise is unbounded, not only would it impart more destructive effects on the protocol, but would also create hindrances towards the successful implementation through the inability to gauge the minimizing unitaries. Furthermore, we also report that the inherent randomness present in the noise would help to counteract the destructive effects of non-Markovianity, thereby highlighting a positive impact of an otherwise undesirable element.}
Our work thus indicates that non-Markovianity and random unitaries used in enhancing the performance of the dense coding protocol have a trade-off relation.

\section*{Acknowledgement}
We acknowledge the support from the Interdisciplinary Cyber-Physical Systems (ICPS) program of the Department of Science and Technology (DST), India, Grant No.: DST/ICPS/QuST/Theme- 1/2019/23.  We acknowledge the use of \href{https://github.com/titaschanda/QIClib}{QIClib} -- a modern C++ library for general-purpose quantum information processing and quantum computing (\url{https://titaschanda.github.io/QIClib}) and cluster computing facility at Harish-Chandra Research Institute.  This research was supported in part by the 'INFOSYS scholarship for senior students'.
\appendix
\section{non-Markovian noisy channels}
\label{subsec:noise_channel}

Noise  acts on the states after encoding at the sender's end occurs. 
In this work, we concentrate on two channels, the dephasing, and the depolarising channels. The non-Markovian versions of the aforementioned channels are parameterized by the non-Markovianity parameter $\alpha$ and the noise strength $p$. The Kraus operators for the dephasing and depolarising channels take the form as \cite{Shrikant_PRA_2018}
    \begin{eqnarray}
		&& K_{I}^{dph} =\sqrt{[1-\alpha p](1-p)} \mathbb{I}, K_{z}^{dph} = \sqrt{[1+\alpha(1-p)]p}\sigma_{z}, \nonumber \\
		\label{eq:deph_channel}\\
		&& K_{I}^{dp} =\sqrt{[1-3\alpha p](1-p)} \mathbb{I},  K_{i}^{dp} = \sqrt{\frac{[1+3\alpha(1-p)]p}{3}}\sigma_{i}. \nonumber\\
		\label{eq:depo_channel}
\end{eqnarray}
Here, $\sigma_i$ with $i = x,y,z$ represent the well-known Pauli matrices and the non-Markovianity parameter $\alpha$ lies between $0$ and $1$. For the dephasing channel, we have $0 \leq p \leq 0.5$ whereas for the depolarising channel $p$ runs from $0$ to $1/3\alpha$. The Markovian limit is recovered by choosing $\alpha = 0$, and a higher value of $\alpha$ corresponds to a higher amount of non-Markovianity in the channel \cite{Daffer_PRA_2004}.


\section{Noisy dense coding involving a Bell state}
\label{sec:bell_dcc}

Let us consider that the Bell state $|\phi^+\rangle_{\mathcal{S}\mathcal{R}} = (1/\sqrt{2})(|00\rangle + |11\rangle)$ is shared between a sender, $\mathcal{S}$ and a receiver, $\mathcal{R}$. Upon encoding, the channel through which $\mathcal{S}$ sends the qubit to $\mathcal{R}$ is either a non-Markovian dephasing channel or a depolarising one. The dense coding capacity is given by 
\textcolor{black}{
\begin{equation}
    C^1_{\text{noise}}(|\phi^+\rangle_{\mathcal{S}\mathcal{R}}) = 2 - S(\Lambda(U^{\min} |\phi^+\rangle_{\mathcal{S}\mathcal{R}} \langle \phi^+| U^{\text{min} \dagger })).
    \label{eq:DCC-1s1r}
\end{equation}
}
For analytical simplicity, we will not consider the unitaries $U^{\min}$ while calculating the capacity. This will not make any qualitative changes in the end result. Note that, the state is said to be dense codeable when $S(\Lambda(|\phi^+\rangle_{\mathcal{S}\mathcal{R}} \langle \phi^+|)) \leq 1$.\\
Let us first consider the depolarising channel. The eigenvalues of $\Lambda(|\phi^+\rangle_{\mathcal{S}\mathcal{R}} \langle \phi^+|)$ are given by $\{x, (1-x)/3, (1-x)/3, (1-x)/3\}$ where $x = (1 - p)(1 - 3 \alpha p)$ when $\alpha > 0$ which reduces to $x = (1-p)$ in the Markovian limit.
It can easily be shown that the entropy term is below $1$ when $p \leq 1/3\alpha$, which is the range of operation for the depolarising channel. By setting $\alpha = 0$, we find that in the case of the Markovian channel, the state remains dense codeable as long as $p \lesssim 0.19$. Now, given the set of eigenvalues of the noisy state, it is evident that the maximum of $S(\Lambda(|\phi^+\rangle_{\mathcal{S}\mathcal{R}} \langle \phi^+|))$ occurs when $x = 0.25$. For the Markovian case, we have $(1 - p) \geq 0.25$ and $p/3 \leq 0.25$. Conversely, when $\alpha > 0$, $(1 - p)(1 - 3 \alpha p) \geq 0.25$ and $p(1 + 3 \alpha (1 - p))/3 \leq 0.25$. Note that if the eigenvalue less than $0.25$ is greater in the non-Markovian case as compared to the Markovian limit, then the last entropy term is also greater when $\alpha \neq 0$. 
We can indeed observe that  $S(\Lambda_{NM}(|\phi^+\rangle_{\mathcal{S}\mathcal{R}} \langle \phi^+|)) \geq S(\Lambda_M(|\phi^+\rangle_{\mathcal{S}\mathcal{R}} \langle \phi^+|))$ and the dense coding capacity for the non-Markovian case is lower than that of the Markovian case when the noisy channel is depolarising in nature.

We now move on to the case of the dephasing channel. The eigenvalues of the noisy resource in this case read as $\{\frac{1}{2}(1 \pm \sqrt{1 + 4 p (p-1)(\alpha (p-1) - 1)(\alpha p -1)})\}$. 
We compare the last  term in the DCC for
Markovian and non-Markovian channels.  Since the eigenvalues are symmetric around $1/2$, we can conclude that $S(\Lambda_{NM}(|\phi^+\rangle_{\mathcal{S}\mathcal{R}} \langle \phi^+|)) \leq S(\Lambda_M(|\phi^+\rangle_{\mathcal{S}\mathcal{R}} \langle \phi^+|))$ when the lower eigenvalue for $\alpha > 0$ is less than the lower eigenvalue for $\alpha = 0$ (which implies that the higher eigenvalue in the non-Markovian case is greater than that in the Markovian limit). Let us define $l$ as the difference between the lower eigenvalues of the Markovian and non-Markovian cases. It can easily be verified that as $\alpha \to 1$ and $p \to 0.5$, $l \geq 0$ since its minimum value is zero and its maximum value $\approx 0.18$. Thus in the limit of high noise strength and high non-Markovian parameter, the entropy term for the non-Markovian channel is lower than that of the corresponding Markovian one. Therefore, the dense coding capacity of the non-Markovian channel is greater than that for $\alpha = 0$. On the other hand, as $p \to 0$,
the dense coding capacity at low noise strengths is higher for the Markovian channel than that of the non-Markovian one. Thus in the case of dephasing noise, non-Markovianity allows for an advantage over the Markovian regime only when the noise strength and the non-Markovian parameter both take high values.\\

\bibliographystyle{apsrev4-1}
	\bibliography{DCCbib}

\end{document}